\documentstyle[12pt,epsf,dina4p]{article}                             %
\newcommand{\figfI}     {and1.epsi}                   
\newcommand{\figfII}    {and2.epsi}                   
\newcommand{\figfIII}   {and3.epsi}                   
\newcommand{\figI}      {scanmu+.eps}                 
\newcommand{\fighI}     {scanmu_high.eps}             
\newcommand{\figII}     {rb_vgl.eps}                  
\newcommand{\fighII}    {rb_vgl_hi.eps}               
\newcommand{\figV}      {mssmbsg1.6_11_sw_znn.eps}    
\newcommand{\figVI}     {mssmbsg35_11_sw_znn.eps}     
\newcommand{\figchiI}   {chi2_314.eps}                
\newcommand{\figchiII}  {chi2_315.eps}                
\newcommand{\figchiIII} {chi2_316.eps}                
\newcommand{\figchiIV}  {chi2_317.eps}                
\newcommand{\mtmh}      {mtmh4.eps}                   
\newcommand{\smdchi}    {smdchi.eps}                  
\newcommand{\sintw}     {sintw.eps}                   
\newcommand{\figochiall}{chi2_25_n.eps}               
\newcommand{\figochi}   {chi2_26_n.eps}               
\newcommand{\figorb}    {Rb_26_n.eps}                 
\newcommand{\figobsg}   {bsg_26_n.eps}                
\newcommand{\figoalf}   {alphas_26_n.eps}             
\newcommand{\figomix}   {phimix_26_n.eps}             
\newcommand{\fighochi}  {chi2_23_n_hi.eps}            
\newcommand{\fighorb}   {Rb_23_n_hi.eps}              
\newcommand{\fighobsg}  {bsg_23_n_hi.eps}             
\newcommand{\fighoalf}  {alphas_23_n_hi.eps}          
\newcommand{\fighomix}  {phimix_23_n_hi.eps}          
\newcommand{\alphas}  {alphas.eps}                    
\newcommand{\bq} {\begin{equation}}
\newcommand{\eq} {\end{equation}}
\newcommand{\nn}   {\nonumber \\}
  
\begin{document}                                                         
\sloppy
\begin{titlepage}
\begin{flushright}
\vspace*{-0.6cm}
\noindent
 \hfill IEKP-KA/96-08   \\
 \hfill KA-TP-18-96\\
  \hfill hep-ph/9609209   \\
 \hfill revised, November 11th, 1996 \\
\end{flushright}
\vspace{0.4cm}
\begin{center} {\Large\bf Global Fits of the SM and \\MSSM 
         to Electroweak Precision Data \\}
\vspace{0.5cm}
{\small {\bf  W.~de Boer$^i$\footnote{Email: wim.de.boer@cern.ch}, 
 A.~Dabelstein$^{iii}$\footnote{E-mail: Andreas.Dabelstein@tu-muenchen.de}, 
 W.~Hollik$^{ii}$\footnote{E-mail: hollik@itpaxp3.physik.uni-karlsruhe.de}, \\
 W.~M\"osle$^{ii}$\footnote{E-mail: wm@itpaxp1.physik.uni-karlsruhe.de}, 
 U.~Schwickerath$^{i}$\footnote{E-mail: Ulrich.Schwickerath@cern.ch},
\\
}\vspace{0.5cm}
{\it i) Inst.\ f\"ur Experimentelle Kernphysik, Univ.\   Karlsruhe, \\}
{\it Postfach 6980, D-76128 Karlsruhe, Germany  \\} 
{\it ii) Inst.\ f\"ur Theoretische Physik, Univ.\   Karlsruhe, \\}
{\it Postfach 6980, D-76128 Karlsruhe, Germany  \\}} 
{\it iii) Inst.\ f\"ur Theoretische Physik, Univ.\   M\"unchen, \\}
{\it James Franck Stra\ss e, D-85747 Garching, Germany}
\end{center}

\vspace{0.5cm}

\begin{center}
{\bf Abstract}\\
\end{center}

A program including all radiative corrections to the Minimal Supersymmetric Standard Model
(MSSM) at the same
level as the radiative corrections to the Standard Model (SM) has been developed and used
to perform global fits to all electroweak data from LEP, SLC and the Tevatron 
and the radiative $b\rightarrow s\gamma$ decay from CLEO.
Values of the strong coupling constant at the $M_Z$ scale and 
$\sin^2\theta_{\overline{MS}}$ are derived, both in the SM and MSSM.

Recent updates on electroweak data, which have been presented at the Warsaw Conference
in summer 1996,
reduce the too high branching ratio of the $Z^0$ boson into b-quarks 
in the SM
from a $3.2 \sigma$ to a $1.8 \sigma$ effect.
In addition, the $b\rightarrow s\gamma$ decay is 30\%  below the SM prediction.
In the MSSM light stops and
light  charginos increase $R_b$ and decrease the  $b\rightarrow s\gamma$ rate, so both observations can be brought into agreement with the MSSM for the 
same region of parameter space.
However, the resulting    $\chi^2$ value for the MSSM fits is only marginally lower
 and in addition,  the splitting in the stop sector has to be  unnaturally high.

\parindent0.0pt
\newpage
\thispagestyle{empty}
\vspace{-10cm}.
\newpage
\end{titlepage}
\pagenumbering{arabic}

\section{Introduction}
Supersymmetry presupposes a symmetry between fermions and bosons, which
can be implemented in the Standard Model (SM) by introducing a fermion for each boson 
and vice versa\cite{rev}.
In this case the problem of quadratic divergent radiative 
corrections to the Higgs boson masses
is solved, since fermions and bosons contribute with an opposite sign 
to the loop corrections.
These new supersymmetric particles (``sparticles'') contribute 
additionally to the radiative corrections
and can influence electroweak precison variables, like e.g.
$R_b=\Gamma_{Z^0\rightarrow b\bar b}/\Gamma_{Z^0\rightarrow hadrons}$ or the penguin mediated
decay $b\rightarrow s\gamma$.
Radiative corrections have been calculated in the Minimal Supersymmetric Standard Model
 (MSSM) to nearly the same level  as in the SM, so an equivalent analysis of 
all electroweak data can be performed both in the SM and MSSM.
In this paper such analysis are described 
using   data from Tevatron \cite{tevatron,top},
LEP and SLC \cite{Lep3,Lep2},
the measurement of
$R_{b\rightarrow s\gamma}=\frac{BR(b\rightarrow s\gamma)}{BR(b\rightarrow ce\bar\nu)}$ from CLEO \cite{cleo}
and limits on the masses of supersymmetric
particles \cite{lim1,lim2,lim4,lim6,rev96}.
For the SM predictions the ZFITTER program\cite{zfitter} was used, while for the MSSM predictions
the SUSYFITTER program\cite{dab/hollik}, which will be discussed below, was used.
In both cases MINUIT\cite{minuit} was used as $\chi^2$ minimizer in order to obtain the optimum
parameter values.
For all SUSY masses well above the electroweak scale one does not expect significant
differences between the SM and the MSSM predictions.

Previous LEP data \cite{Lep2} showed a too high value of $R_b$ (3.2$\sigma$) and a too low
value of $R_c$ (2$\sigma$).
It has been shown by several 
groups\cite{yel1}\nocite{boufi,chan2,ell1,garc3,kan1,kan2,garcia}-\cite{garcia2}
that it is possible to increase $R_b$
in the MSSM  with light charginos,  top squarks or Higgses, which yield large 
positive contributions to the $Zb\bar b$ vertex because of the large Yukawa couplings 
of the third generation.
The first two generations are not affected by such corrections, so 
 no modifications in $R_c$ can  be
obtained.
Recent updates of electroweak data,  presented at 
the Warsaw Conference\cite{Lep3},
 show no  deviation of $R_c$  anymore
 and a value of $R_b$ which is $1.8\sigma$ above the SM value.
 In addition, the experimental  value for
$R_{b\rightarrow s\gamma}=\frac{BR(b\rightarrow s\gamma)}{BR(b\rightarrow ce\bar\nu)}$ 
from CLEO \cite{cleo}
 equals $(2.32\pm0.67)\cdot10^{-4}$, which is
 about 30\%  below the SM prediction
 after  taking the  ca. 10\% increase in the prediction
 by the next-to-leading-log contributions into account\cite{misiak}.

In the MSSM an increase in $R_b$ can  
cause a decrease in the  
$b\rightarrow s\gamma$ rate, since both cases involve similar diagrams usually with an 
opposite sign, 
 e.g. the $\tilde{t}-\tilde{\overline{t}}-\chi^\pm$ 
vertex corrections in $Z^0\rightarrow b\overline{b}$ and the 
$\tilde{t}-\chi^\pm$ loop corrections in $b\rightarrow s\gamma$ have an opposite sign. 
Both, the too high  value of $R_b$ and the too low value of $b\rightarrow s\gamma$
  prefer  a chargino mass around 
90 GeV and a stop  mass around 50 GeV or
alternatively  light Higgses around 50 GeV, so both observations 
agree with the MSSM for the same
region of parameter space, albeit an unnatural one, as will be shown.
%
%

\section{Radiative Corrections in the MSSM}

At the $Z$ boson resonance two classes of precision observables are available:
\begin{itemize}
\item[a)] inclusive quantities:
  \begin{itemize}
   \item[$\bullet$] the partial leptonic and hadronic decay width $\Gamma_{f 
                  \bar{f}}$,
   \item[$\bullet$] the total decay width $\Gamma_Z$,        
   \item[$\bullet$] the hadronic peak cross section $\sigma_h$, 
   \item[$\bullet$] the ratio of the hadronic to the electronic decay
                  width of the $Z$ boson: $R_h$, 
   \item[$\bullet$] the ratio of the partial decay width for $Z\rightarrow c\bar{c} \,
                  ( b \bar{b} )$ to the hadronic width, 
                  $R_c$, $R_b$.
  \end{itemize}
\item[b)] asymmetries and the corresponding mixing angles:
  \begin{itemize}
   \item[$\bullet$] the \it forward-backward \rm asymmetries $A_{FB}^f$,
   \item[$\bullet$] the \it left-right \rm  asymmetries $A_{LR}^f$,
   \item[$\bullet$] the $\tau$ polarization $P_\tau$,
  \end{itemize}
\end{itemize}
This set of
precision observables together with the $R_{b\rightarrow s\gamma}$ rate\cite{cleo}, 
the $W$ - and the top mass 
is convenient for a numerical analysis of the 
supersymmetric parameter space.
In the following the observables defined above are expressed
with the help  of effective couplings.

\subsection{The effective $Z$-$f$-$f$ couplings}\label{zeff}                                     

The coupling of the $Z$ boson to fermions $f$ can be expressed by effective
vector and axial vector coupling constants $v_{eff}^f, \, a_{eff}^f$ in terms of the
NC vertex:
\begin{equation}
J_{NC}^\mu = \frac{e}{2 s_W c_W} \, \gamma^\mu \, ( v_{eff}^f - a_{eff}^f \gamma_5 ) \ ,
\end{equation}
where the convention is introduced : $c_W^2 = \cos^2 \theta_W = 1 - s_W^2 = M_W^2 / 
M_Z^2$ \cite{sirlin}.
Input parameters are the $\mu$ decay constant $G_\mu = 1.166392 \times 10^{-5}$
GeV$^{-2}$, $\alpha_{EM} = 1/137.036$ and the mass of the $Z^0$ boson $M_Z = 91.1884$
GeV. The mass of the $W$ boson is related to these input parameters through:
\begin{eqnarray}\label{gmudr}                                                   
{G_{\mu}\over\sqrt{2}} = {\pi\alpha_{EM}\over 2 s^2_W M^2_W} \cdot
\frac{1}{
1 - \Delta r_{MSSM} \left(\alpha_{EM},M_W,M_Z,m_t,...\right)} \ ,
\end{eqnarray} 
where the complete MSSM one-loop contributions are parameterized by the
quantity $\Delta r_{MSSM}$\cite{sola}.
Leading higher order SM corrections\cite{bhp,fleischer} to the 
quantity $\Delta r$ are included in the calculation.
\smallskip 

The effective couplings $v_{eff}^f, \, a_{eff}^f$ can be written as:
\begin{eqnarray}
v_{eff}^f & = & \sqrt{Z_Z} \, (v^f + \Delta v^f + Z_M Q_f) \nonumber \\
a_{eff}^f & = & \sqrt{Z_Z} \, (a^f + \Delta a^f) \ .
\end{eqnarray}
$v^f$ and $a^f$ are the tree-level vector and axial vector couplings:
\begin{equation}
v^f = I_3^f - 2 Q_f s_W^2 \ , \ a^f =   I_3^f.
\end{equation}
$Z_Z$, $Z_M$ are given in eq. (\ref{ZZ}).
The complete MSSM one-loop contributions of the non-universal 
finite vector and 
axial vector couplings $\Delta v^f$, $\Delta a^f$ have been 
calculated\cite{dab/hollik},
together with the leading two-loop Standard Model 
contributions\cite{bhp,fleischer,Chet/Kw}.
They are derived in the 't Hooft-Feynman gauge and in the
on-shell renormalization scheme\cite{bhs}.
Fig.~ 1 shows the MSSM one-loop $Z \rightarrow f \bar{f}$ 
vertex correction diagrams.
\vspace{2cm}
\begin{figure}[ht]
 \begin{center}
  \leavevmode
  \epsfxsize=16cm
  \epsffile{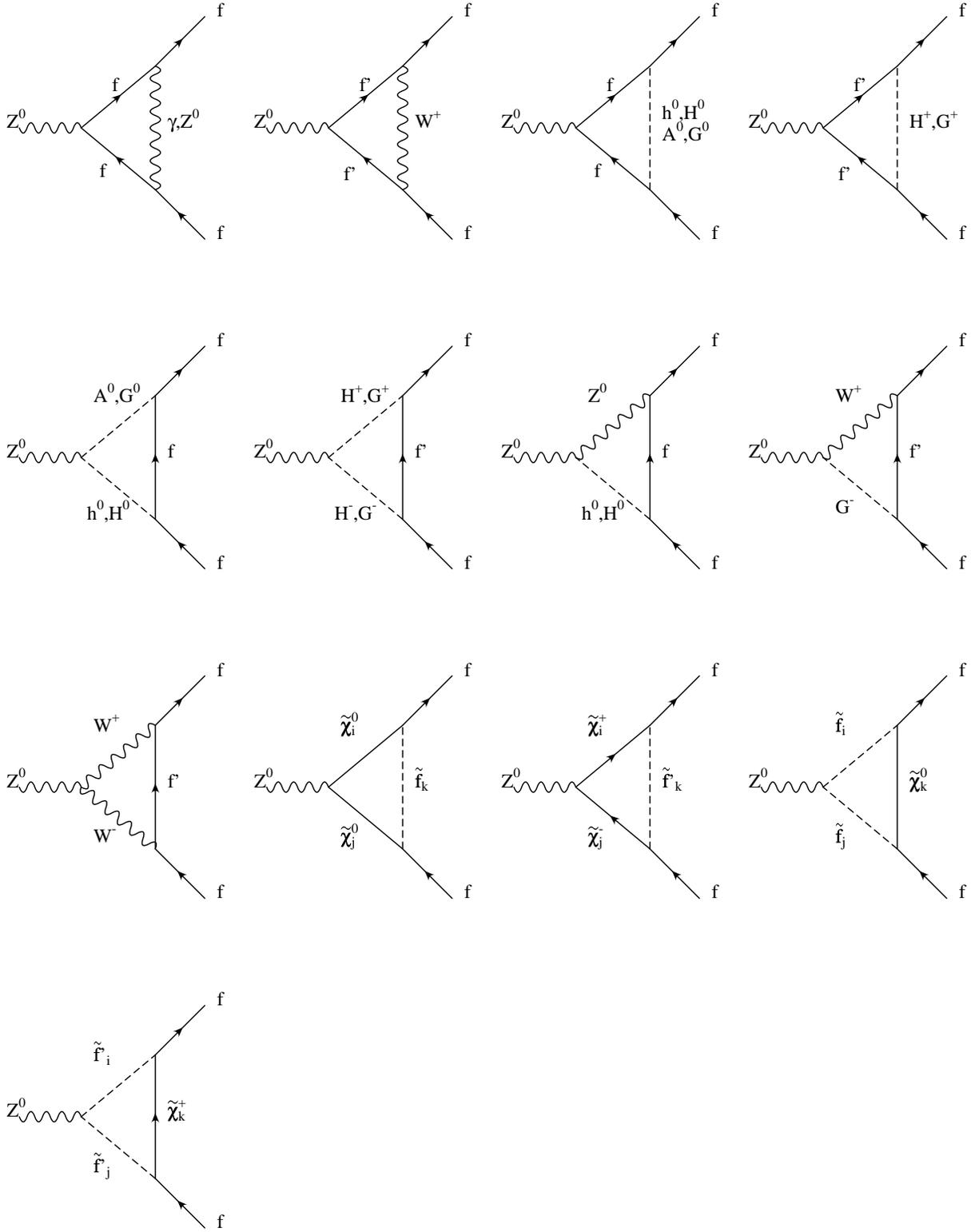}
\end{center}
\caption{\label{\figfIII} MSSM one-loop $Z \rightarrow f \bar{f}$ vertex correction 
diagrams. $i,j,k = 1,..,2 (4)$ are chargino ($\tilde\chi^{\pm}$), neutralino
($\tilde\chi^{0}$)
and sfermions ($\tilde{f}$) indices, while h, H, are the scalar Higgs bosons, A is the
pseudoscalar Higgs boson, and G are Goldstone bosons. 
No particle permutations are shown.
}
\end{figure}
%
%
\begin{figure}
 \begin{center}
  \leavevmode
  \epsfxsize=12cm
  \epsffile{\figfII}
\end{center}
\caption{\label{\figfII}
MSSM  fermion self-energies. For notation see fig.\protect\ref{\figfIII}.}

\begin{center}
  \leavevmode
  \epsfxsize=16cm
  \epsffile{\figfI}
\end{center}
\caption{\label{\figfI}$Z$ boson wave function renormalization. For notation see fig.\protect\ref{\figfIII}.}
\end{figure}
\vfill
%
\clearpage
\noindent
The non-universal contributions  can be written  in the following way:
\begin{eqnarray}
 \Delta v_f & = & F_V^{SM} + \Delta F_V , \nonumber \\
 \Delta a_f & = & F_A^{SM} + \Delta F_A . \nonumber
\end{eqnarray}
The Standard Model form factors $F_{V,A}^{SM}$ corresponding to the
diagrams of figs. 1 and 2 can be found
e.g in refs.~\cite{bhs,bhp}.
The diagrams with a virtual photon are listed for completeness in the figures.
They are not part of the effective weak couplings but are treated separately
in the QED corrections, together with real photon bremsstrahlung.
The non-standard contributions are summarized by
\begin{eqnarray}
\Delta F_V & = &  \sum_i F_V^{(i)} 
                + v_f \delta Z_V^f + a_f \delta Z_A^f \nonumber \\
\Delta F_A & = &  \sum_i F_A^{(i)} 
                + v_f \delta Z_A^f + a_f \delta Z_V^f \nonumber 
\end{eqnarray}
where the sum extends over the diagrams of fig. 1 with internal
charged and neutral Higgs bosons, charginos, neutralinos and scalar 
fermions. Each diagram contributes 
$$
    F_V^{(i)} \gamma_{\mu} - F_A^{(i)} \gamma_{\mu}\gamma_5
$$
to the $Zff$-vertex.
The self-energy diagrams of fig. 2 with internal neutral Higgs,
chargino, neutralino and sfermion lines determine the field renormalization
constants
\begin{eqnarray}
 \delta Z_V^f & = & -\Sigma_V(m_f^2) -
                  2m_f^2 [ \Sigma'_V(m_f^2) + \Sigma'_S(m_f^2) ]
                  \nonumber \\
 \delta Z_A^f & = & \Sigma_A(m_f^2) 
\end{eqnarray}
with the scalar functions $\Sigma_{V,A,S}$ in the decomposition of the 
fermion self-energy according to
 $$
  \Sigma = \not{p} \Sigma_V + \not{p} \gamma_5 \Sigma_A + m_f \Sigma_S . $$
The contributions from the Higgs sector are given explicitly in 
ref.~\cite{dghk}. For the genuine SUSY diagrams, the
couplings for  charginos, neutralinos and sfermions are taken 
from \cite{hunter}, together with the diagonalization matrices given in 
the next section.

\medskip 
The universal propagator corrections from the 
 finite $Z$ boson wave function renormalization $Z_Z$ and the $\gamma Z$
mixing $Z_M$ are derived from the $(\gamma, \, Z)$ propagator matrix. The 
inverse matrix is:
\begin{equation}
(\Delta_{\mu \nu})^{-1} = i g_{\mu \nu} \, \left(
 \begin{array}{ll}
 k^2 + \hat{\Sigma}_\gamma (k^2) &  \hat{\Sigma}_{\gamma Z} (k^2) \\
 \hat{\Sigma}_{\gamma Z} (k^2)   &  k^2 - M_Z^2 + \hat{\Sigma}_Z (k^2)
 \end{array} \right) \ ,
\label{gzmatrix}
\end{equation}
where $\hat{\Sigma}_\gamma$, $\hat{\Sigma}_{Z}$, $\hat{\Sigma}_{\gamma Z}$
are the renormalized self energies and mixing.
They are obtained by summing the loop diagrams, shown symbolically in fig. \ref{\figfI}, 
and the counter terms and can be found in ref.~\cite{dab/hollik}.
\smallskip

The entries in the $(\gamma, \, Z)$ propagator matrix:
\begin{equation}
\Delta_{\mu \nu} = - i g_{\mu \nu} \left( \begin{array}{ll}
\Delta_\gamma     &  \Delta_{\gamma Z} \\
\Delta_{\gamma Z} &  \Delta_Z \end{array} \right) \ ,
\end{equation}
are given by:
\begin{eqnarray}
\Delta_\gamma (k^2) & = & \frac{1}{k^2 +  \hat{ \Sigma}_\gamma (k^2) -
 \frac{\hat{ \Sigma}_{\gamma Z}^2 (k^2)}{k^2 - M_Z^2 + \hat{ \Sigma}_Z (k^2)} }
 \nonumber \\
 \Delta_Z (k^2) & = & \frac{1}{k^2 - M_Z^2 + \hat{ \Sigma}_Z (k^2) -
 \frac{\hat{ \Sigma}_{\gamma Z}^2 (k^2)}{k^2  + \hat{ \Sigma}_\gamma (k^2)} }
 \nonumber \\
\Delta_{\gamma Z} (k^2) & = & -\frac{\hat{ \Sigma}_{\gamma Z} (k^2)}{
 [ k^2  + \hat{ \Sigma}_\gamma (k^2)] \, 
 [ k^2 - M_Z^2 + \hat{ \Sigma}_Z (k^2) ] - \hat{ \Sigma}_{\gamma Z}^2 (k^2) } \ .
\label{zprop}
\end{eqnarray}
The renormalization condition to define the mass of the $Z$ boson is given
by the pole of the propagator matrix (eq.~\ref{gzmatrix}). 
The pole $k^2 = M_Z^2$ is the solution of the equation: 
\begin{equation}
 \cal R\rm e \, [ \, ( M_Z^2   + \hat{ \Sigma}_\gamma (M_Z^2) \, ) \, 
 \hat{ \Sigma}_Z (k^2)  - \hat{ \Sigma}_{\gamma Z}^2 (M_Z^2) \, ] = 0 \ .
\end{equation}
Eq. (\ref{zprop}) yields the wave function renormalization $Z_Z$ and mixing $Z_M$:
\begin{eqnarray}
Z_Z & = & Res_{M_Z} \Delta_Z = \left. \frac{1}{1 +
        \cal R\rm e \hat{ \Sigma}_Z' (k^2) -
        \cal R\rm e
   \left( \frac{ \hat{ \Sigma}_{\gamma Z}^2 (k^2)}{k^2 + \hat{ \Sigma}_\gamma (k^2)
   } \right)'  } \ \right| _{k^2 = M_Z^2} \nonumber \\
Z_M & = & - \frac{\hat{ \Sigma}_{\gamma Z} (M_Z^2)}{M_Z^2 + 
 \hat{ \Sigma}_\gamma (M_Z^2) } \ .
\label{ZZ}
\end{eqnarray}

\subsection{ $Z$ boson observables}               

\noindent
The fermionic $Z$ boson partial decay widths $\Gamma_{f \bar{f}}$ can be 
written as follows:
\begin{itemize}
\item[1)] $f \ne b$:

\begin{eqnarray}
 \Gamma_{f \bar{f}} & = & \frac{N_C \, 
 \sqrt{2} G_{\mu} M_Z^3}{12\pi} (1-\Delta r_{MSSM}) \,
 \left[ (v_{eff}^f)^2 
  + (a_{eff}^f)^2 (1 - \frac{6 m_f^2}{M_Z^2} ) \right] \cdot \nonumber \\
  & & \cdot (1 + \frac{3 \alpha_{EM}}{4 \pi} Q_f^2) \, (1 + \delta_{QCD}^f ) \ ,
\end{eqnarray}
where 
\begin{equation}
 \delta_{QCD}^f = \left\{ \begin{array}{ll}
  0 & ,f = \mbox{leptons} \nonumber \\
  \frac{\alpha_s}{\pi} + 1.405 (\frac{\alpha_s}{\pi})^2 - 12.8
  (\frac{\alpha_s}{\pi})^3
   -\frac{Q_f^2}{4}\frac{\alpha\alpha_s}{\pi^2} & ,f = \mbox{quarks}
 \end{array} \right. \ .
\label{deqc}
\end{equation}
\item[2)] $f = b$:
\begin{eqnarray}
 \Gamma_{b \bar{b}} & = & \frac{N_C \, 
 \sqrt{2} G_{\mu} M_Z^3}{12\pi} (1-\Delta r_{MSSM}) \,
 \left[ (v_{eff}^b)^2 + (a_{eff}^b)^2  \right] \nonumber \\
   &  & 
 \cdot (1 + \frac{3 \alpha_{EM}}{4 \pi} Q_b^2) \, (1 + \delta_{QCD}^b ) \,
 + \Delta \Gamma_{b \bar{b}} \ . \nonumber \\
\end{eqnarray}
In $\Delta \Gamma_{b \bar{b}}$ the $b$ quark specific finite mass terms 
with QCD corrections \cite{Chet/Kw} are included.
 $\delta_{QCD}^b$ is given in eq. (\ref{deqc}).
\end{itemize}
The total decay width $\Gamma_Z$ is the sum of the contributions from leptons and quarks: \\
 \begin{equation}
   \Gamma_Z = \sum_{f} \Gamma_{f \bar{f}} \ .
 \end{equation}
 In the following $\Gamma_{had} = \sum_{q} \Gamma_{q \bar{q}}$ is the hadronic decay
 width of the $Z$ boson.
\medskip

\noindent
The hadronic peak cross section is defined as
\begin{equation}
 \sigma_h = \frac{12 \pi}{M_Z^2} \frac{\Gamma_{ee} \Gamma_{had}}{\Gamma_Z^2} \ .
\end{equation}
The ratio of the hadronic to the electronic decay width is defined as
\begin{equation}
 R_e = \frac{\Gamma_{had}}{\Gamma_{e e}} \ .
\end{equation}
The ratio of the partial decay width for
$Z \rightarrow b \bar{b} \, (c \bar{c})$ to
the total hadronic decay width is given by
\begin{equation}
 R_{b (c)} = \frac{\Gamma_{b \bar{b} (c \bar{c}) }}{\Gamma_{had}} \ .
\end{equation}
\medskip

The following quantities and observables depend on the ratio of the vector to
axial vector coupling. The effective flavour dependent weak mixing angle can
be written as
\begin{equation}
 \sin^2 \theta_{eff}^f = \frac{1}{4 | Q_f |} \, \left( 1 - \frac{v_{eff}^f}{a_{eff}^f} 
 \right) \ .
\end{equation}
The \it left-right \rm asymmetries are given by

\begin{equation}
 A_{LR}^f = {\cal A}^{\rm f}  = \frac{2 \, v_{eff}^f / a_{eff}^f}{1 + ( v_{eff}^f /
 a_{eff}^f )^2 } \ ,
\end{equation}
while the \it forward-backward \rm asymmetries can be written as
\begin{equation}
 A_{FB}^f = \frac{3}{4} \, \cal A^{\rm e} \rm \, \cal A^{\rm f} \rm \ .
\end{equation}

\section{The MSSM}

\subsection{Higgs sector}

%
The MSSM has two Higgs doublets:
\bq H_1= \left(\begin{array}{c}v_1 +\frac{1}{\sqrt{2}}
\left(H^0\cos\alpha-h^0\sin\alpha
+iA^0\sin\beta-iG^0\sin\beta\right)\\
H^-\sin\beta-G^-\cos\beta\end{array}\right)\eq
\bq H_2= \left(\begin{array}{c}
H^+\cos\beta+G^+\sin\beta\\
v_2 +\frac{1}{\sqrt{2}}
\left(H^0\sin\alpha+h^0\cos\alpha
+iA^0\cos\beta+iG^0\sin\beta\right)
\end{array}\right)\eq
Here $H ,h $ and $A $ represent
the neutral Higgs bosons, while
the $G$'s represent the Goldstone fields, which
correspond to  the longitudinal
polarization components of the heavy gauge bosons. 
The imaginary  and real sectors do not mix, since
they have different CP-eigenvalues;
$\alpha$ and $\beta$
are the mixing angles in these different sectors, 
so one is left with 2 neutral CP-even
Higgs bosons $H^0$ and $h^0$,
 one CP-odd neutral Higgs bosons $A^0$, and two 
charged Higgs bosons.

Their masses are  completely determined by the ratio of
the vacuum expectation values 
of $\tan\beta=v_2/v_1$ and the pseudoscalar mass $M_A$, together with the 
radiative corrections. The latter ones are taken into account in terms of
the effective potential approximation with the leading terms $\sim m_t^4$,
including the mixing in the scalar top system \cite{ellisetal}.
In this way, the coupling constants of the various Higgs particles
to gauge bosons and fermions can be taken over from \cite{hunter}
substituting only the scalar mixing angle $\alpha$ by the improved 
effective mixing angle which is obtained from the diagonalization of the
scalar mass matrix, discussed in the next section.

\subsection{Sfermion sector}

The physical masses of squarks and sleptons are given by the eigenvalues
of the $2 \times 2$ mass matrix:
\begin{equation}
 \cal{M}_{\tilde{f}}^{\rm 2} \rm = \left( \begin{array}{ll}
 M_{\tilde{Q}}^2 + m_f^2 + M_Z^2 (I_3^f - Q_f s_W^2) \cos 2 \beta &
 m_f (A_f + \mu \{ \cot \beta , \tan \beta \} ) \\
 m_f (A_f + \mu \{ \cot \beta , \tan \beta \} ) &
 M_{\{\tilde{U},\tilde{D}\}}^2 + m_f^2 + M_Z^2 Q_f s_W^2 \cos 2 \beta
 \end{array} \right) \ ,
\label{sqmatrix}
\end{equation}
with SUSY soft breaking parameters $M_{\tilde{Q}}$,
$M_{\tilde{U}}$, $M_{\tilde{D}}$, $A_f$, and the mass parameter $\mu$ from 
the Higgs sector \cite{rev}.
It is convenient to use  the following notation for the 
off-diagonal entries in eq. (\ref{sqmatrix}):
\begin{equation}
A_f' = A_f + \mu \{ \cot \beta , \tan \beta \} \ .
\label{glaprime}
\end{equation}
Scalar neutrinos appear only as left-handed mass eigenstates.
Up and down type sfermions in (\ref{sqmatrix}) are distinguished by
setting $f=u~(d)$ and select $\cot\beta$ ($\tan\beta$) in the curly brackets.
Since the non-diagonal terms are proportional to $m_f$, it seems
natural to assume unmixed sfermions for the lepton and quark case
except for the scalar top sector.
The $\tilde{t}$ mass matrix is diagonalized by a rotation matrix with
a mixing angle $\Phi_{mix}$. 
Instead of $M_{\tilde{Q}}$, $M_{\tilde{U}}$, $M_{\tilde{D}}$,  $A_t'$
for the $\tilde{b}$, $\tilde{t}$ system
the physical squark masses
$m_{\tilde{b}_L}, m_{\tilde{b}_R}$, $m_{\tilde{t}_2}$ can be used
together with 
$A_t'$ or, alternatively, the stop mixing angle $\Phi_{mix}$.
For simplicity we assume $m_{\tilde{b}_L}=m_{\tilde{b}_R}$, and 
$\tilde{u}$, $\tilde{d}$, $\tilde{c}$, $\tilde{s}$
to have masses equal to the 
$\tilde{b}$ squark mass.

\smallskip
A possible mass splitting between
 $\tilde{b}_L$-$\tilde{t}_L$  yields a contribution to the 
$\rho$-parameter\footnote{The superscript
 in $\Delta \rho^0$ indicates that no left-right mixing is present.} 
$\rho = 1+ \Delta\rho^0$ 
in terms of \cite{sola}:
\begin{equation} 
\Delta \rho_{\tilde{b}-\tilde{t}}^0 = \frac{3 \alpha_{EM}}{16 \pi s_W^2 M_W^2} \,
 \left(m_{\tilde{b}_L}^2 + m_{\tilde{t}_L}^2 - 2 \frac{m_{\tilde{b}_L}^2 
 m_{\tilde{t}_L}^2}{m_{\tilde{b}_L}^2 - m_{\tilde{t}_L}^2} \log \frac{
 m_{\tilde{b}_L}^2}{m_{\tilde{t}_L}^2} \right) . 
\end{equation}
As a universal loop contribution, it enters the quantity
\begin{equation} 
 \Delta r \simeq \Delta \alpha_{EM} -
  \frac{c_W^2}{s_W^2} \, \Delta \rho^0 \, + ... \nonumber    
\end{equation}
and all the $Z$ boson widths 
$$  \Gamma_{f\bar{f}} \sim 1 + \Delta\rho^0 + \cdots $$
and is thus significantly constrained by the data on $M_W$ and the
leptonic widths.

\subsection{Chargino/Neutralino sector}
\noindent
The chargino (neutralino) masses and the mixing angles in the gaugino
couplings are calculated from $\mu$ and the soft breaking parameters $M_1$, $M_2$ in
the chargino (neutralino) mass matrix\cite{hunter}. 

The chargino $2 \times 2$ mass matrix is given by
\begin{equation}\label{charmat}
  \cal{M}_{\rm \tilde{\chi}^\pm} \rm = \left( \begin{array}{ll}
    M_2 & M_W \sqrt{2} \sin \beta \\ M_W \sqrt{2} \cos \beta & - \mu \\
    \end{array}  \right) \ .
\end{equation} \\
%
The physical chargino mass states $\tilde{\chi}^{\pm}_i$ 
are the rotated
wino and charged Higgsino states:
\begin{eqnarray}
\tilde{\chi}^+_i & = & V_{ij} \psi^+_j   \nonumber \\
\tilde{\chi}^-_i & = & U_{ij} \psi^-_j  \ ; \ i,j = 1,2  \ .
\end{eqnarray}
$V_{ij}$ and $U_{ij}$ are unitary chargino mixing matrices obtained from
the diagonalization of the mass matrix eq.~\ref{charmat}:
\begin{equation}
\rm U^* \cal{M}_{\rm \tilde{\chi}^\pm} \rm  V^{-1} = diag(m_{\tilde{\chi}^\pm_1},m_{\tilde{\chi}^\pm_2}) \ .
\end{equation}
\smallskip \par
The neutralino $4 \times 4 $ mass matrix can be written as:   
\begin{equation}\label{neutmat}
 \cal{M}_{\rm \tilde{\chi}^0}   = \left( \begin{array}{cccc}
 M_1 & 0 & - M_Z \sin \theta_W \cos \beta & M_Z \sin \theta_W \sin \beta \\
 0 & M_2 & M_Z \cos \theta_W \cos \beta & - M_Z \cos \theta_W \sin \beta \\
- M_Z \sin \theta_W \cos \beta & M_Z \cos \theta_W \cos \beta & 0 &  \mu \\
 M_Z \sin \theta_W \sin \beta & - M_Z \cos \theta_W \sin \beta & \mu & 0
 \\  \end{array} \right) 
\end{equation}
where the diagonalization can be obtained by the unitary matrix $N_{ij}$: 
\begin{equation}
  \rm N^* \cal{M}_{\rm \tilde{\chi}^0} \rm N^{-1}  = diag(
  m_{\tilde{\chi}_i^0}) \ .
\end{equation}

\smallskip \noindent
The elements $U_{ij}$, $V_{ij}$, $N_{ij}$ of the diagonalization
matrices  enter the couplings of the 
charginos, neutralinos and sfermions to fermions and gauge bosons, as
explicitly given in ref.~\cite{hunter}. Note that our sign convention on the
parameter $\mu$ is opposite to that of ref.~\cite{hunter}.

\begin{table}[ht]
\begin{center}
\begin{tabular}{|c|c|}
\hline
\hline
\multicolumn{2}{|c|}{experimental limits}\\   
\hline
\hline
$m_{\tilde{\chi}^{\pm}_{1,2}}$ & $>$ 65~GeV \\
\hline
$m_{\tilde{\chi}^{0}_1}$ & $>$ 13~GeV \\
$m_{\tilde{\chi}^{0}_2}$ & $>$ 35~GeV \\
$m_{\tilde{\chi}^{0}_{3,4}}$ & $>$ 60~GeV \\
$\Gamma_{Z\rightarrow neutralinos}$ & $<2$~MeV\\
\hline
$m_{\tilde{t}_{1,2}}$ & $>$ 48~GeV \\
\hline
$m_h$,$m_H$,$m_A$,$m_{H^\pm}$  & $>$ 50~GeV\\
\hline
\end{tabular}
\end{center}
\caption{\label{limits}
Mass limits assumed for the optimized fits.}
\end{table}

\begin{table}[th]
\begin{center}
\begin{tabular}{|l|c||cc|cc|cc|}
\hline
\hline
Symbol & measurement & \multicolumn{6}{|c|}{ best fit observables}\\
\hline
\hline
 & & \multicolumn{2}{|c|}{SM} & \multicolumn{4}{|c|}{MSSM}\\
\hline
$\tan\beta$ and pull        &                    &    & pull  & 1.6 & pull & 35 & pull\\
\hline
\hline
\hline
$M_Z$ [GeV]                 & 91.1863  $\pm$ 0.0020   &  91.1861  &   0.08 & 91.1863 &    -  &  91.1863 &  - \\
$\Gamma_{Z}$[GeV]           &  2.4946  $\pm$ 0.0027   &   2.4958  &  -0.45 &  2.4946 & -0.00 &   2.4940 & 0.22\\
$\sigma_h$ [nb]             & 41.508   $\pm$ 0.056    &  41.468   &   0.72 & 41.461  &  0.84 &  41.449  & 1.05\\
$R_l$                       & 20.778   $\pm$ 0.029    &  20.755   &   0.80 & 20.769  &  0.32 &  20.772  & 0.22\\
$A_{FB}^l$                  &  0.0174  $\pm$ 0.0010   &   0.0160  &   1.41 &  0.0162 &  1.20 &   0.0162 & 1.22\\
$R_b$                       &  0.2178  $\pm$ 0.0011   &   0.2158  &   1.75 &  0.2174 &  0.38 &   0.2168 & 0.92\\
$R_c$                       &  0.1715  $\pm$ 0.0056   &   0.1722  &  -0.13 &  0.1707 &  0.14 &   0.1708 & 0.12\\
$A_{FB}^b$                  &  0.0979  $\pm$ 0.0023   &   0.1022  &  -1.87 &  0.1031 & -2.26 &   0.1031 &-2.24\\
$A_{FB}^c$                  &  0.0735 $\pm$ 0.0048   &   0.0731  &   0.10 &   0.0736 & -0.02 &    0.0736 & 0.01\\
${\cal A}_b$                &  0.863 $\pm$ 0.049      &   0.933   &  -1.45 &  0.9353 & -1.49 &   0.9356 & -1.50\\
${\cal A}_c$                &  0.625  $\pm$ 0.084     &   0.667   &  -0.50 &  0.6678 & -0.51 &   0.668    & -0.51\\
\hline                                                                                           
${\cal A}_{\tau}$           &  0.1401 $\pm$ 0.0067     & 0.1460   &  -0.88  & 0.1470  & -1.03 &  0.1466  &-0.97\\  
${\cal A}_e$                &  0.1382 $\pm$ 0.0076    & 0.1460    &  -1.03  & 0.1470  & -1.16  &  0.1469  &-1.14 \\
$\sin^2\theta_{eff}^{lept}\langle Q_{FB}\rangle$ &  0.2320  $\pm$ 0.0010& 0.2316 &  0.35  &0.2315   &  0.48 & 0.2315  & 0.46 \\ 
$M_W$ [GeV]                 &  80.356  $\pm$ 0.125    & 80.355  &  0.01  & 80.403  &  -0.38  &  80.428 & -0.58 \\
$1-M_W^2/M_Z^2$             &  0.2244 $\pm$ 0.0042    &  0.2235 & 0.23   &  0.2225  &  0.45 & 0.2220  &  0.56\\
$m_t$ [GeV]                 &  175    $\pm$ 6.        & 172.0 & 0.50       &172.5  &  0.42   &172.0  & 0.50 \\
$\sin^2\theta_{eff}^{lept}(A_{LR})$ &  0.23061$\pm$ 0.00047 &  0.2316  &  -2.21 &0.2315  &  -1.94  & 0.2315  & -1.97\\
$R_{b\rightarrow s\gamma}$/$10^{-4}$ &  2.32    $\pm$ 0.67 $\pm$0.5& 3.1  &   -0.86 & 2.43  &  -0.12    & 2.33        & 0.0       \\
$1/\alpha(M_Z)$             &  128.896 $\pm$ 0.09      & 128.905 & -0.10 &   128.89     &  -     & 128.89  &   -    \\ 
\hline                    
\end{tabular}             
\caption[]{\label{fitresults} Measurements\cite{Lep3}
and the predicted results of the fits with minimum $\chi^2$.\\
The pulls are defined by (measurement - predicted value) / error of the measurement.
The second error for $R_{b\rightarrow s\gamma}$ has been added to take care of the
uncertainty by the renormalization scale used for the calculation of that quantity.
For the MSSM fits $M_Z$ and $1/\alpha(M_Z)$ were taken as fixed parameters,
because
their uncertainties are negligible compared to  uncertainties arising from the
soft breaking parameters.  Leaving them free does not change the results.
}
\end{center} 
\end{table}

\begin{table}[h]
\protect\vspace{-1.cm}
\begin{center}
\begin{tabular}{|c||r|r|}
\hline
\hline
\multicolumn{3}{|c|}{ Fitted SUSY parameters and masses}\\   
\hline
\hline
Symbol &\makebox[3.0cm]{\bf{$\tan\beta$=1.6}}&\makebox[3.0cm]{\bf{$\tan\beta$=35}}\\
\hline
\hline
\vspace{-0.09cm}
 $m_t$[GeV]                      &172            &172 \\
 $\alpha_s$                      &0.116     &0.1190  \\
 $M_2$[GeV]                      &113             & - \\
 $\mu$[GeV]                      &60             &111 \\
 $m_{\tilde t_2}$[GeV]           &48             &187 \\
 $\phi_{mix}$                    &-0.18     &0.04 \\
 $m_{A}$[GeV]                    &        -                    &50 \\
\hline                                        
\hline                                        
\multicolumn{3}{|c|}{ Particle Spectrum}\\    
\hline                                        
\hline                                        
 $m_{\tilde t_1}$[GeV]          &  \multicolumn{2}{c|}{$\approx 1$~TeV}\\
\cline{2-3} 
 $m_{\tilde t_2}$[GeV]          &   48      &187\\
\cline{2-3} 
 $m_{\tilde q}$[GeV]            & \multicolumn{2}{c|}{{1~TeV}}\\
 $m_{\tilde l} $[GeV]            & \multicolumn{2}{c|}{{0.5~TeV}}\\
\cline{2-3} 
\hline                                         
 $ m_{\tilde{\chi}_1^\pm}$[GeV]          & 149  &1504\\
 $ m_{\tilde{\chi}_2^\pm}$[GeV]          & 84   &111\\
\hline                                        
 $ m_{\tilde{\chi}_1^0}$[GeV]            & 54    &107\\
 $ m_{\tilde{\chi}_2^0}$[GeV]            & 64    &114\\
 $ m_{\tilde{\chi}_3^0}$[GeV]            & 100   &722\\
 $ m_{\tilde{\chi}_4^0}$[GeV]            & 150   &1504\\
\hline                                        
 $ m_{h}$[GeV]                           & 109    &50\\
 $ m_{H}$[GeV]                   &  $\approx 1.5$~TeV   &112 \\
 $ m_{A}$[GeV]                   &  $1.5$~TeV           &50 \\
 $ m_{H^\pm}$[GeV]               &  $\approx 1.5$~TeV    &123\\
\cline{2-3} 
\hline                                                       
 ${\chi^2}/d.o.f.$               & 16.6/12   &18.1/12\\
\hline                                                        
 Probability                     & 17\%      & 11\% \\
\hline                                                       
\hline                                                       
\end{tabular}
\end{center}
\caption[]{\label{bestfit}
Values of the fitted parameters (upper part) and
corresponding mass spectrum (lower part).
On the right hand side the results of the optimization for high $\tan\beta$ are given.
The dashes indicate irrelevant parameters which were chosen high.
}    
\end{table}

\section{Results}


The experimental limits included in the fit are summarized in table \ref{limits}
\cite{lim1,lim2,lim4,lim6,rev96}. They have not been updated with the latest
results, since in practice only $R_{b\rightarrow s\gamma}$ and $R_b$ require light
SUSY particles, but their deviation from the SM has become so small, that
they do not require sparticles below the experimental limits.
The experimental observables  used in the fits  are shown
in the first column of table \ref{fitresults}.
The calculation of the total decay width of the Z boson into neutralinos is based on
reference \cite{zwid1}, the calculation of the ratio $R_{b\rightarrow s\gamma}$ on reference
\cite{bsgamma}. The next-to-leading-log calculations increase the SM prediction for
$R_{b\rightarrow s\gamma}$ by about 10$\%$ to (3.2$\pm$0.5)10$^{-4}
$ \cite{misiak}. This higher order
contribution was taken into account in the first order calculation by choosing the renormalization
scale $\mu=0.7\cdot m_b$.

As discussed in section \ref{zeff}, $R_b$ depends on stop -, chargino- and 
Higgs masses.
First the behaviour of the chargino masses as a function of the SUSY parameters $\mu$ and
$M_2$ is discussed, then
the $R_b$ dependence 
on  the relevant SUSY masses is studied, both  
 for   the low and  high $\tan\beta$ scenarios. These  studies are followed by
the best solutions in the SM and MSSM.

\subsection{Chargino Masses}

\begin{figure*}
 \begin{center}
  \protect\vspace{-2.5cm}
  \leavevmode
  \epsfxsize=9.0cm
  \epsffile{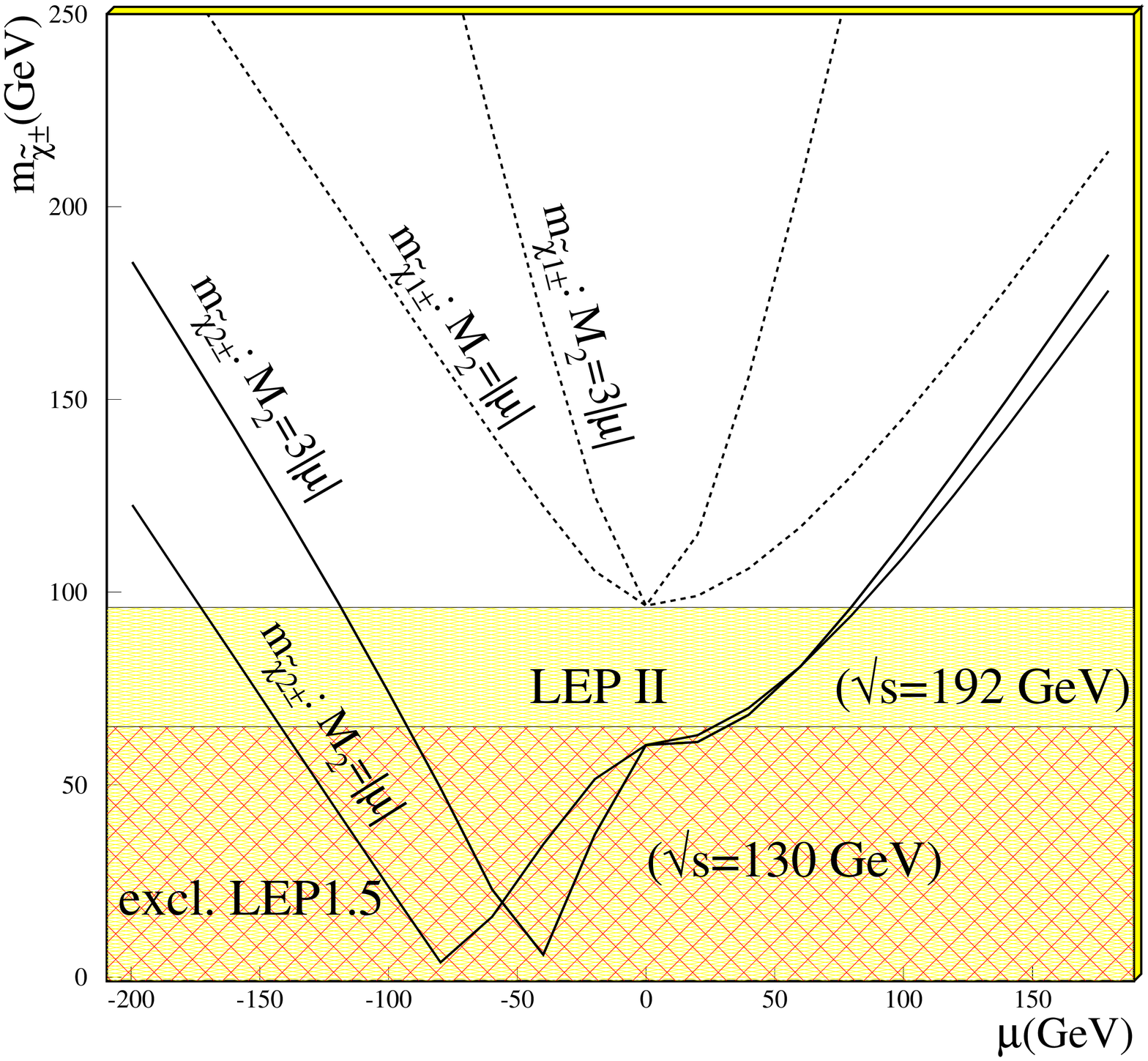}
\end{center}
\protect\vspace{-1.cm}
\caption{\label{\figI} 
Dependence of the chargino masses (solid line=lightest one) on the parameter
$\mu$ for $M_2=|\mu|$ and $M_2=3|\mu|$ 
for $\tan\beta=1.6$, $\alpha_s\approx 0.117$ and $m_{\tilde{t}_2}\approx 60~$GeV.  The
shaded regions indicate chargino masses less than $65$~GeV which are
excluded by $LEP~1.5$ and chargino masses less than $96$~GeV, which is the
discovery reach for LEP II.
Note that 
two light charginos are easier  obtained, if $\mu\approx M_2$. }
\begin{center}
  \protect\vspace{-0.5cm}
  \leavevmode
  \epsfxsize=9.0cm
  \epsffile{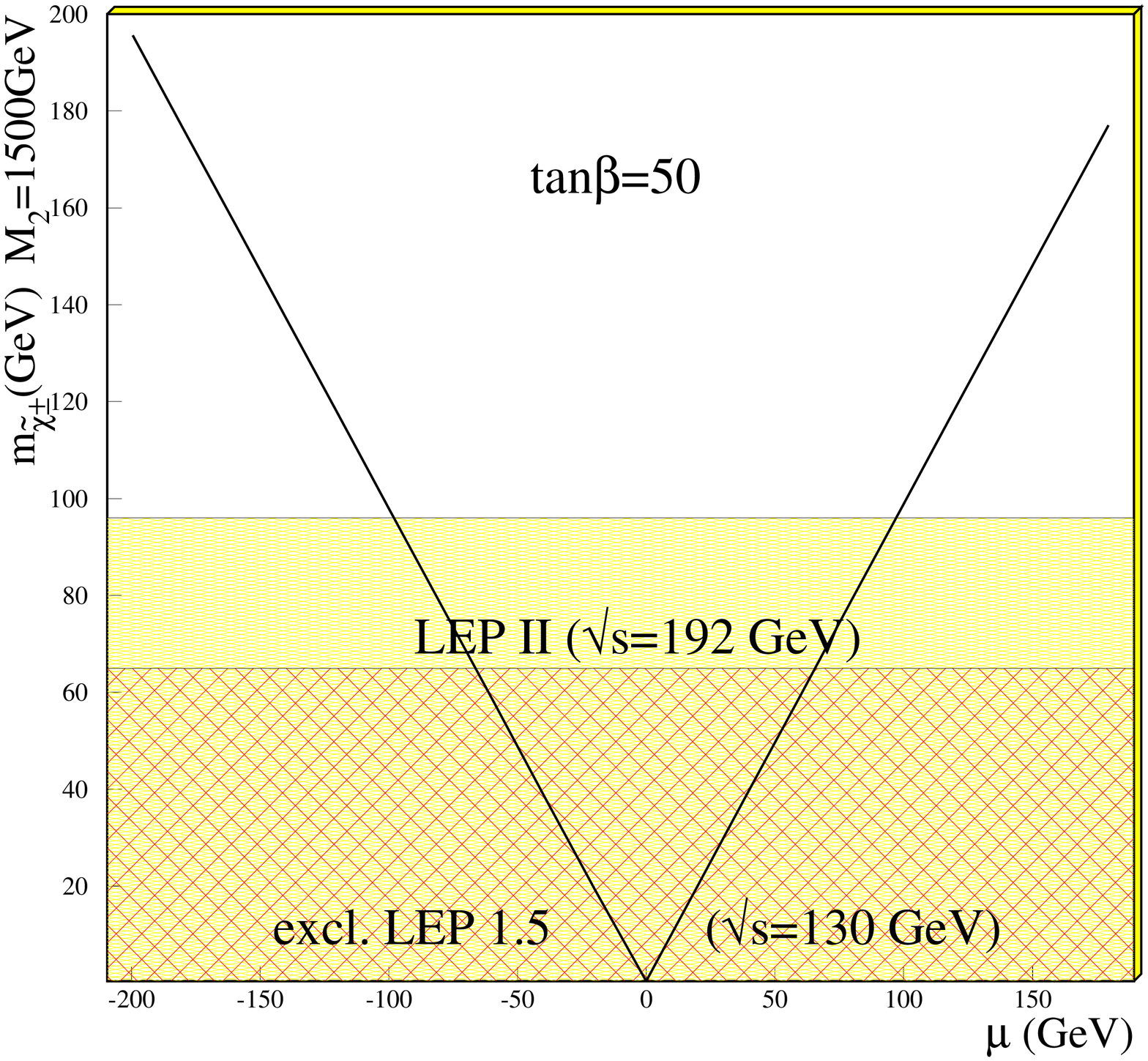}
\end{center}
\protect\vspace{-0.5cm}
\caption[]{\label{\fighI} 
Dependence of the lightest chargino mass  on
the parameter $\mu$ for $\tan\beta=35$ and $M_2$=1500~GeV.
 The shaded regions as in fig. \ref{\figI}.
The heavy chargino
mass is close to  $M_2=1500$ GeV (not indicated). 
}
\end{figure*}

\begin{figure*}
 \begin{center}
  \leavevmode
  \epsfxsize=10cm
  \epsffile{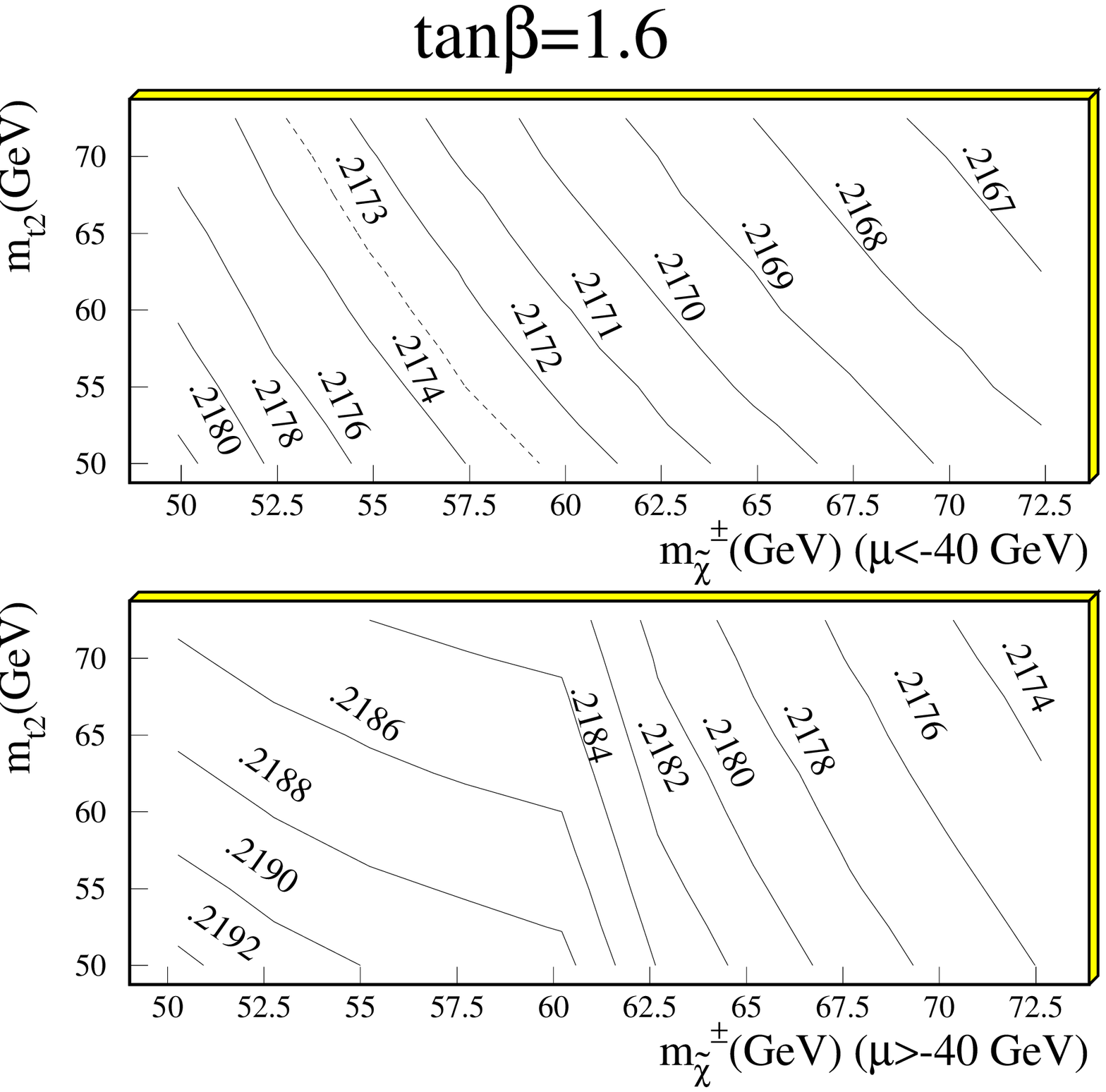}
 \end{center}
\caption{\label{\figII} $R_b$ in the light stop versus light chargino plane with
 $M_2=3|\mu|$ and $\tan\beta=1.6$.
The upper part shows the solution with $\mu<-40~$GeV, in the
lower part the one with $\mu>-40~$GeV
is displayed. In the latter solution
quite high values for $R_b$ are possible, as can be seen in the figure.
The dashed line in the upper plot indicates the old $2\sigma$ lower limit of $R_b$.
Recent updates of electroweak data yield $R_b=0.2178\pm0.0011$.
}
 \begin{center}
  \leavevmode
  \epsfxsize=10cm
  \epsffile{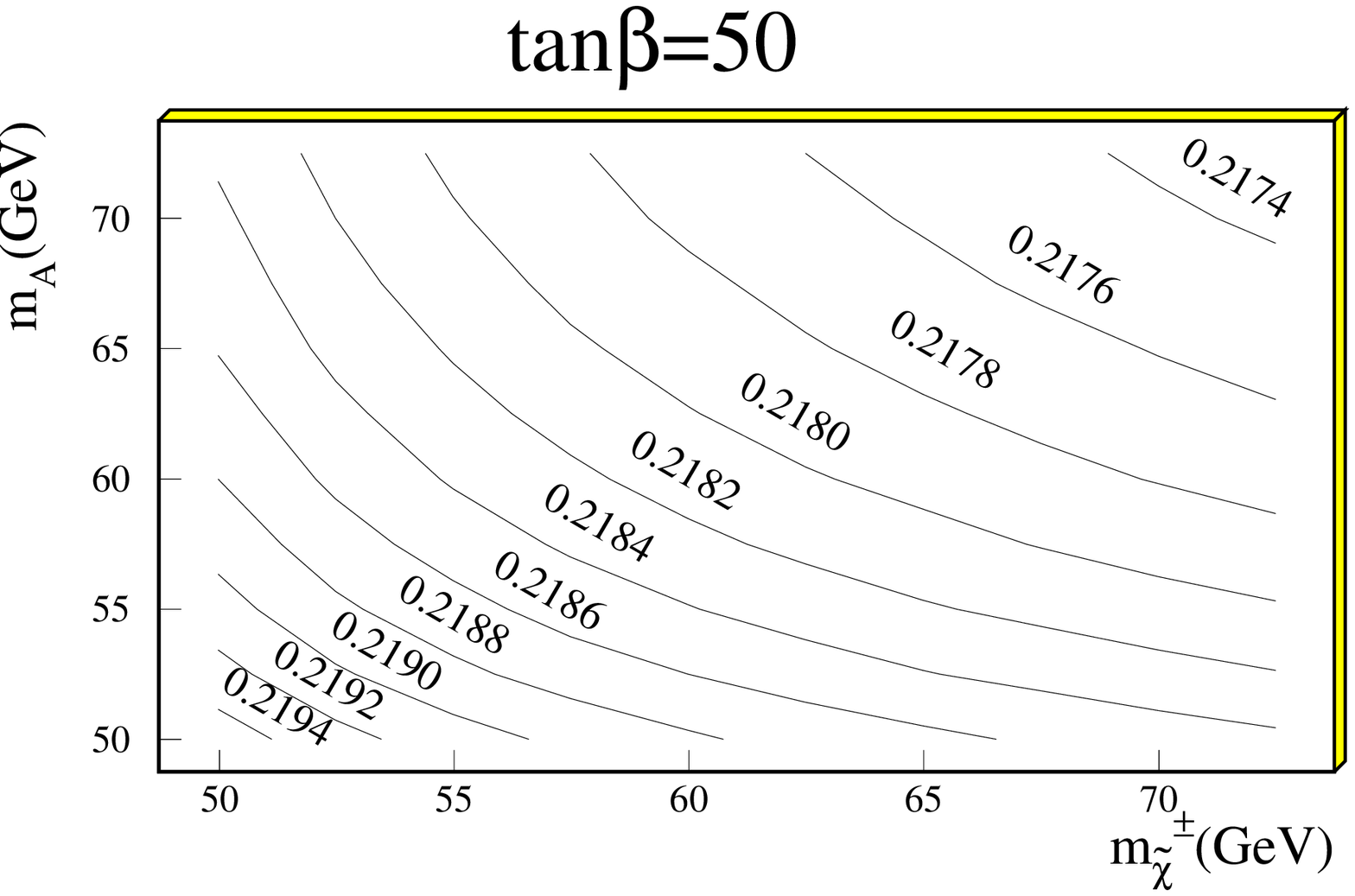}
 \end{center}
\caption{\label{\fighII} $R_b$ in the $m_A$ versus light chargino plane with
  $M_2=3|\mu|$ for the high $\tan\beta$ solution. $\mu$ was chosen positive
  here. In this case choosing the opposite sign for $\mu$ doesn't change $R_b$.
} 
\end{figure*}

In order to explain an enhanced value of $R_b$ in the data,
the MSSM needs a light right handed stop and light  chargino
 (low $\tan\beta$ scenario),
or a light pseudoscalar Higgs $A$ (high $\tan\beta$ scenario) 
\cite{yel1}\nocite{boufi,chan2,ell1,garc3,kan1,kan2,garcia}-\cite{garcia2}.
A higgsino-like chargino can be obtained for a low value of the parameter $\mu$ in the
mass matrix (eq. \ref{charmat}).
Figs.~\ref{\figI}
and ~\ref{\fighI} show the dependence of the chargino masses on the parameter $\mu$.
For high $\tan\beta$, 
$m_{\tilde\chi_{2}}$ is
almost symmetric around $\mu=0$, whereas for low $\tan\beta$
this dependence is more complicated, as can be seen from fig.~\ref{\figI}.
For $M_2=3|\mu|$ the light chargino mass passes zero  at
$\mu=-40$ GeV, so the following low $\tan\beta$ plots were made for
$\mu>-40$ GeV and $\mu\le-40$ GeV.
The asymmetric structure of fig.~\ref{\figI} is reflected in
the contours of constant $R_b$ in the $m_{\tilde\chi_{2}}$ versus light scalar
top $m_{\tilde{t}_2}$ plane (see fig.~\ref{\figII}).
Values of
$R_b$ up to 0.2194 are possible (see figs.~\ref{\figII} and \ref{\fighII}),
although these special regions of the parameter space are already
experimentally excluded by the lower limits on sparticle masses.

\begin{figure*}
\begin{minipage}[t]{8cm}
\leavevmode
\epsfxsize = 7cm
\epsffile{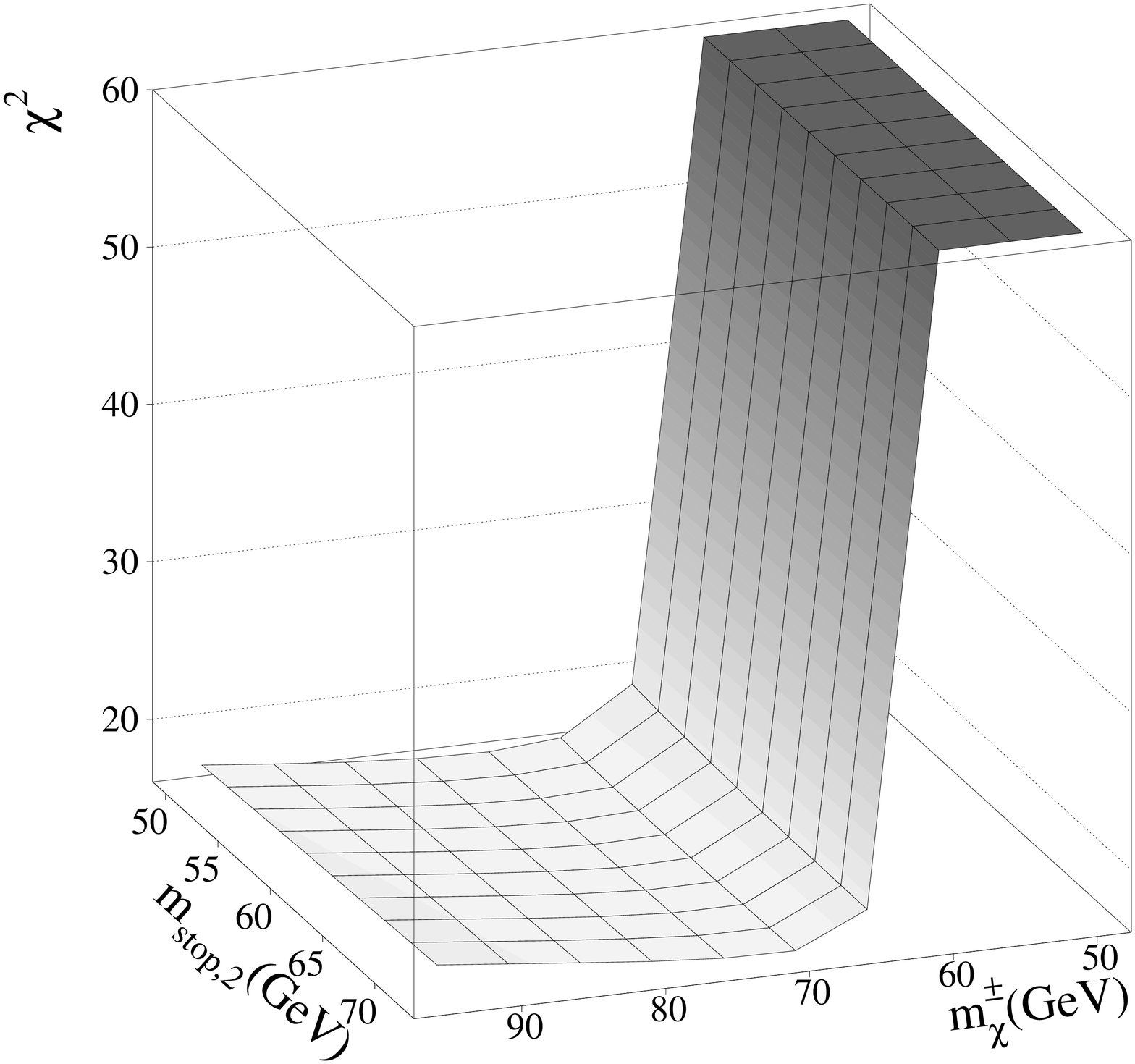}
\end{minipage}
\hfill
\begin{minipage}[t]{8cm}
\epsfxsize = 7cm
\epsffile{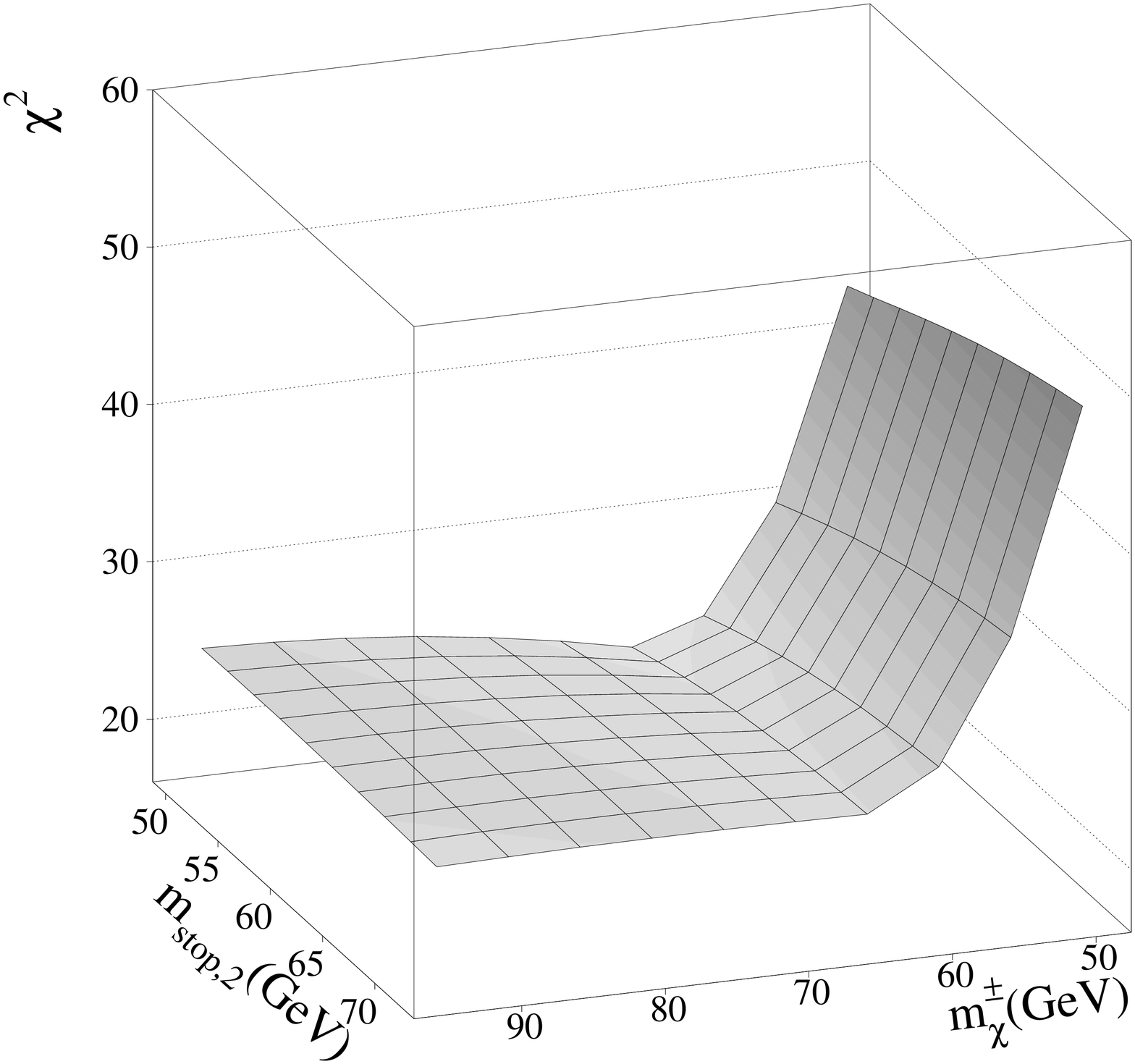}
\end{minipage}
\caption{\label{\figchiI} Dependence of the absolute $\chi^2$ for $\mu>-40~$GeV
(left side) and  $\mu<-40~$GeV (right side), using $M_2=3|\mu|$.
No optimization of parameters was performed, but they were fixed to
values near the minimum.}
\end{figure*}

\begin{figure*}
\begin{minipage}[b]{8cm}
\epsfxsize = 7cm
\epsffile{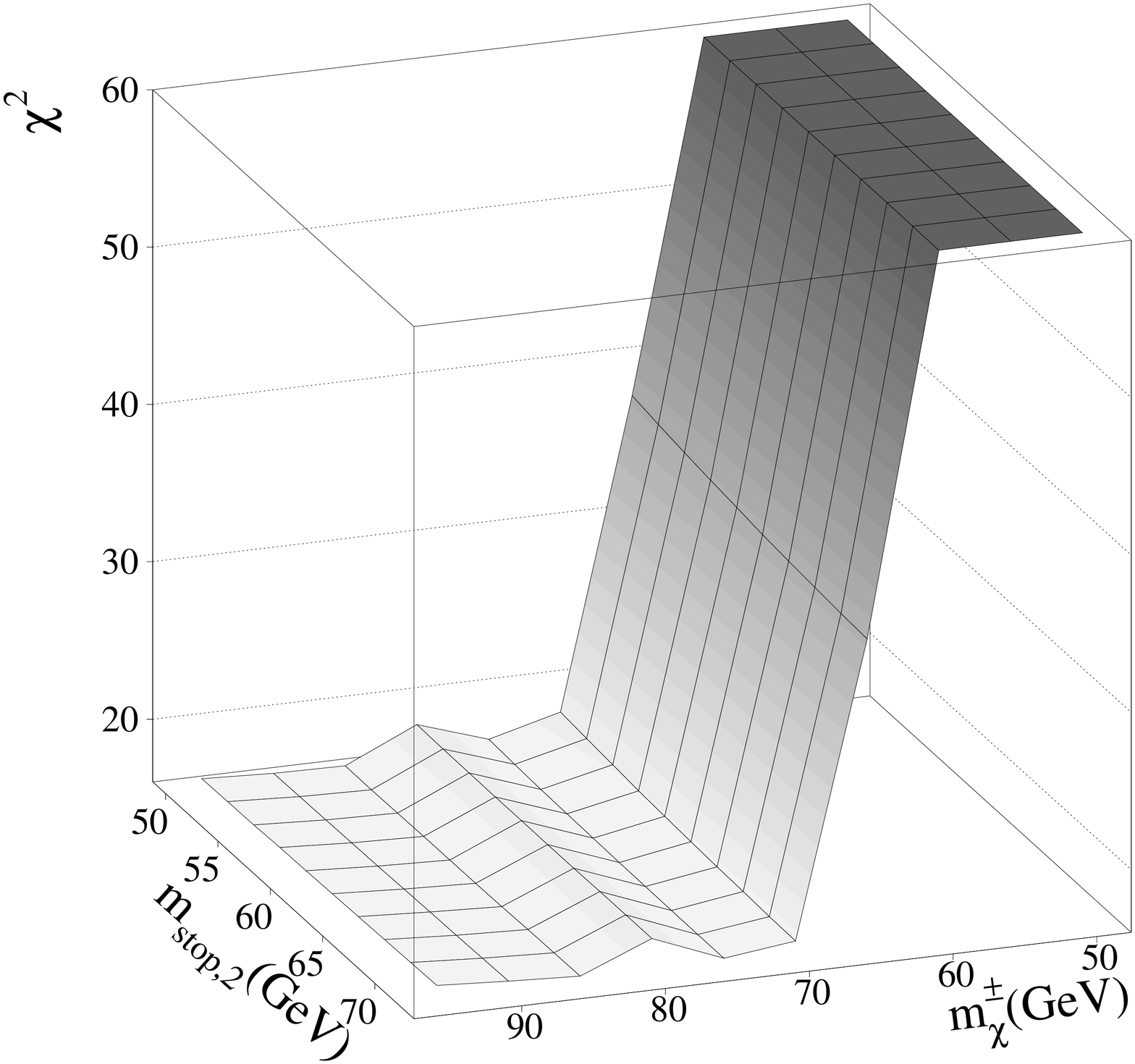}
\end{minipage}
\hfill
\begin{minipage}[b]{8cm}
\epsfxsize = 7cm
\epsffile{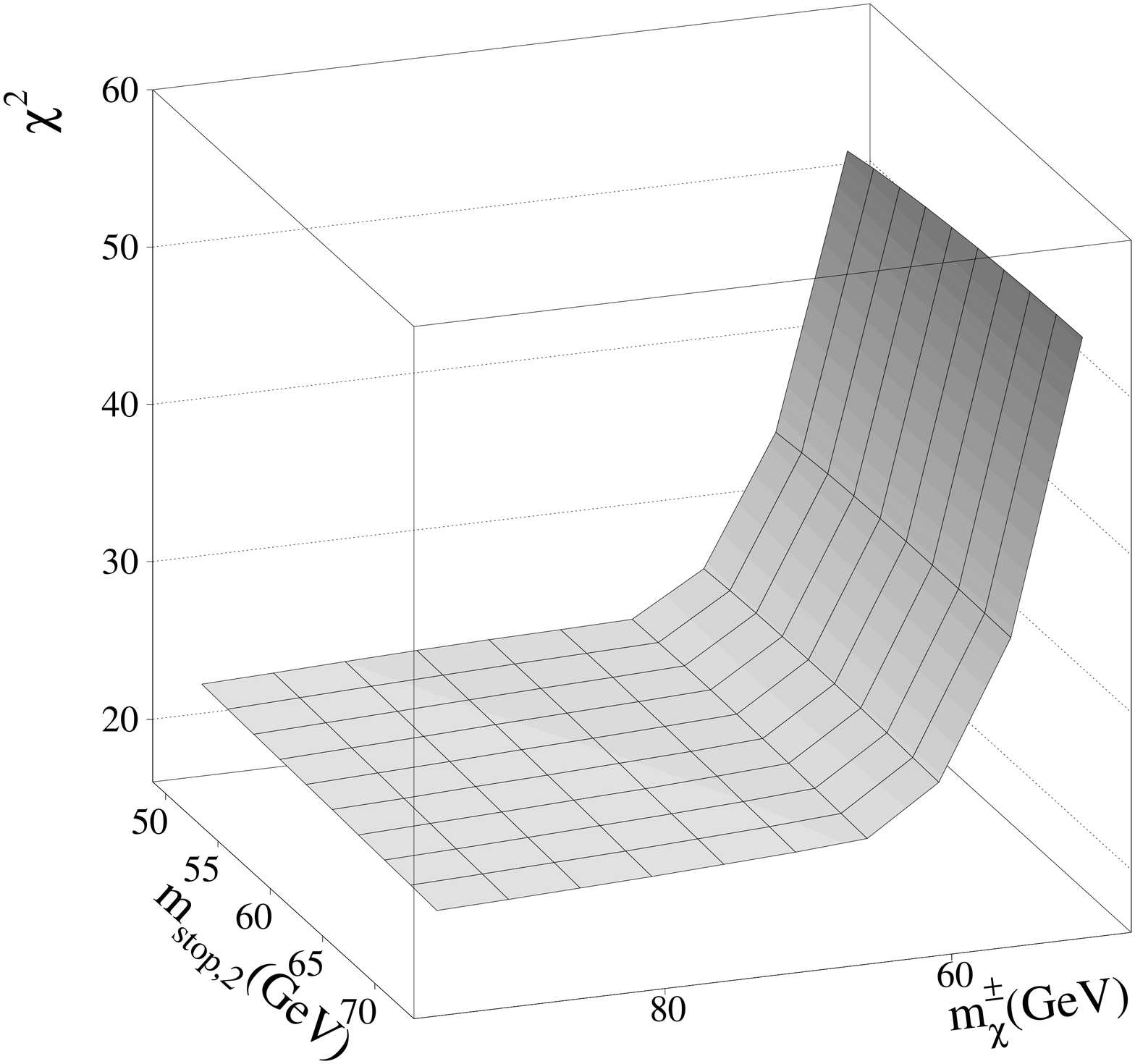}
\end{minipage}
\caption{\label{\figchiIII} The same as \protect\ref{\figchiI}, 
but for $M_2=|\mu|$.
}
\end{figure*}

For $M_2=3|\mu|$ the lightest chargino is mostly higgsino-like, while
the heavier one is gaugino like. 
Mixing the charginos more by taking 
$M_2=|\mu|$ does not change these results very much, as can be seen from
a comparison of the $\chi^2$ distributions in
fig.~\ref{\figchiI} ($M_2=3|\mu|$) and fig.~\ref{\figchiIII} ($M_2=|\mu|$).
The small increase of the $\chi^2$ at
chargino masses around 80~GeV in the left hand part of fig.~\ref{\figchiIII} is due to
neutralino threshold singularities, for which an additional $\chi^2$  contribution
has been added,
when the sum of two neutralino masses is close to the $Z^0$ mass.
The sharp increase of the $\chi^2$ function at low chargino masses is due to experimental
limits on chargino, neutralino and stop 
masses from LEP 1.5~\cite{lim1,lim2,lim4}.

\subsection{Low $\tan\beta$ scenario}

\begin{figure*}
 \begin{center}
  \leavevmode
  \epsfxsize=15cm
  \epsffile{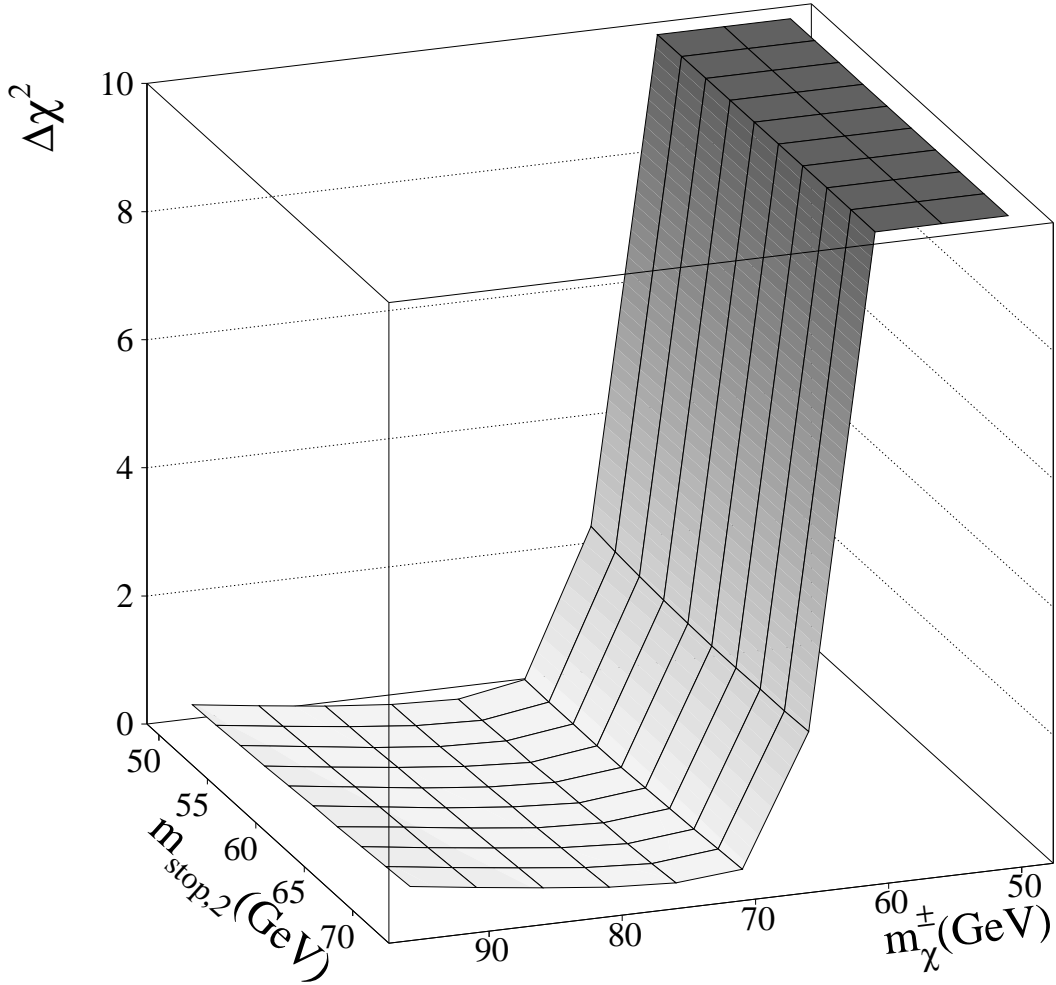}
 \end{center}
\caption{\label{\figochi}
The $\Delta\chi^2$ in the  light stop and light
chargino plane for $\tan\beta=1.6$. At each point of the grid an
optimization of $m_t$, $\alpha_s$
and the stop mixing angle
$\phi_{mix}$ was performed with $\mu > -40$ ~GeV and $M_2=3|\mu|$.
} 
\end{figure*}

\begin{figure*}
 \begin{center}
  \leavevmode
  \epsfxsize=15cm
  \epsffile{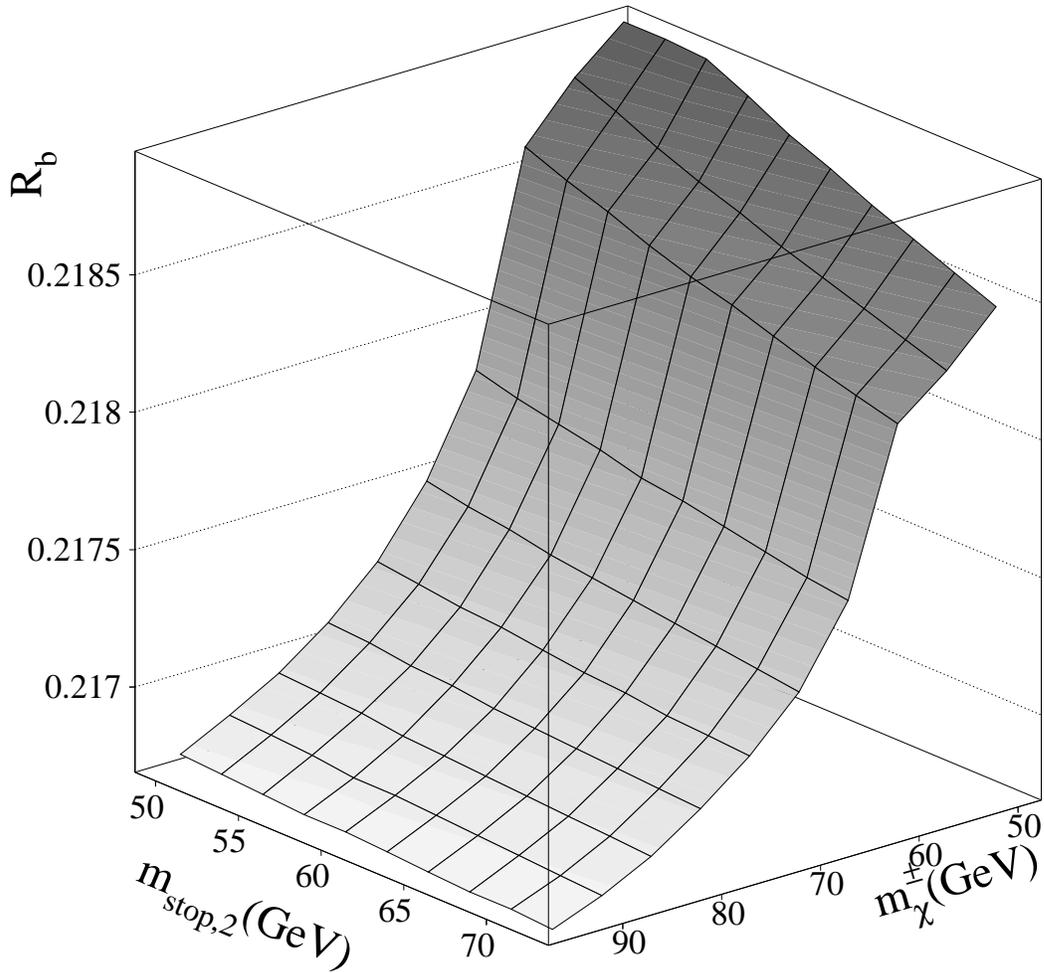}
 \end{center}
\caption{\label{\figorb}
$R_b$ in the  light stop and light
chargino plane. Optimization as in  fig.~\protect\ref{\figochi}.
} 
\end{figure*}

\begin{figure*}
 \begin{center}
  \leavevmode
  \epsfxsize=15cm
  \epsffile{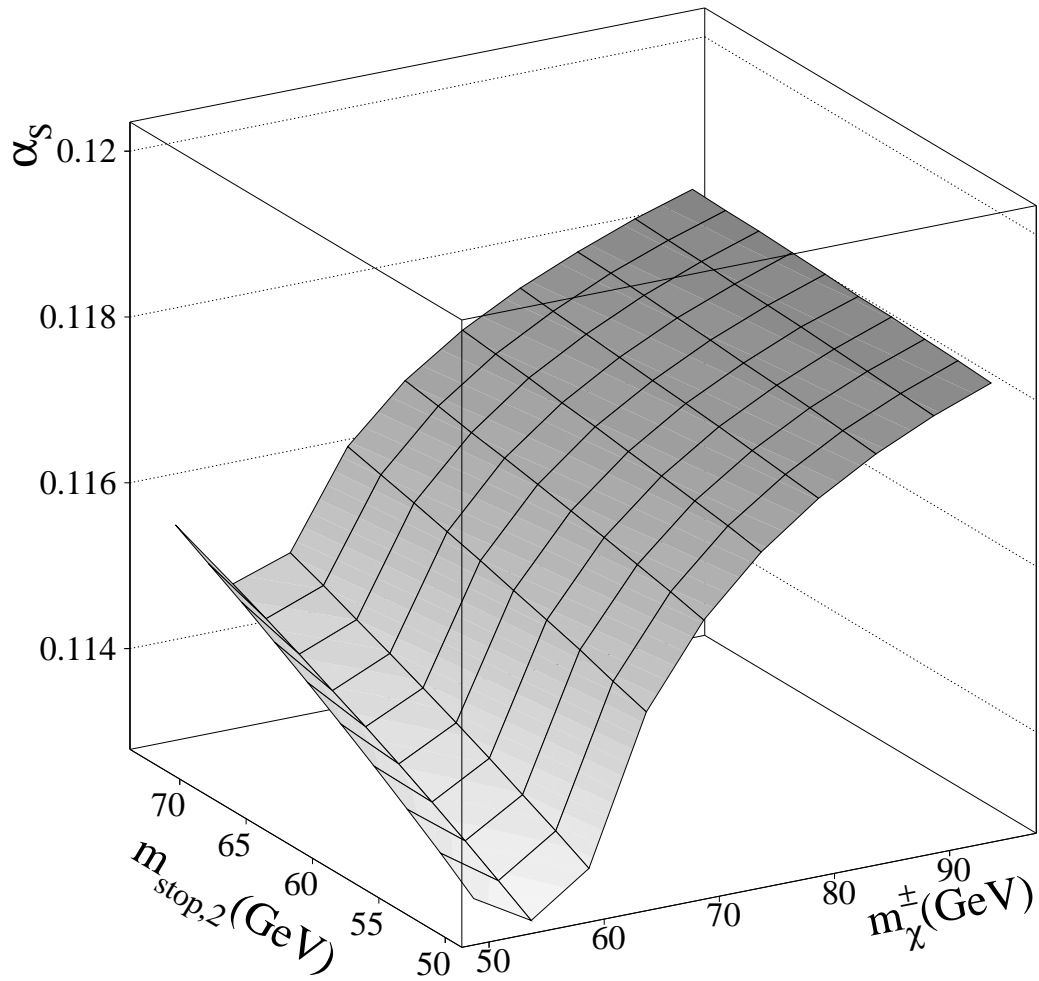}
 \end{center}
\caption{\label{\figoalf}
$\alpha_s$ in the  light stop and light
chargino plane. Optimization as in  fig.~\protect\ref{\figochi}.
} 
\end{figure*}

\begin{figure*}
 \begin{center}
  \leavevmode
  \epsfxsize=15cm
  \epsffile{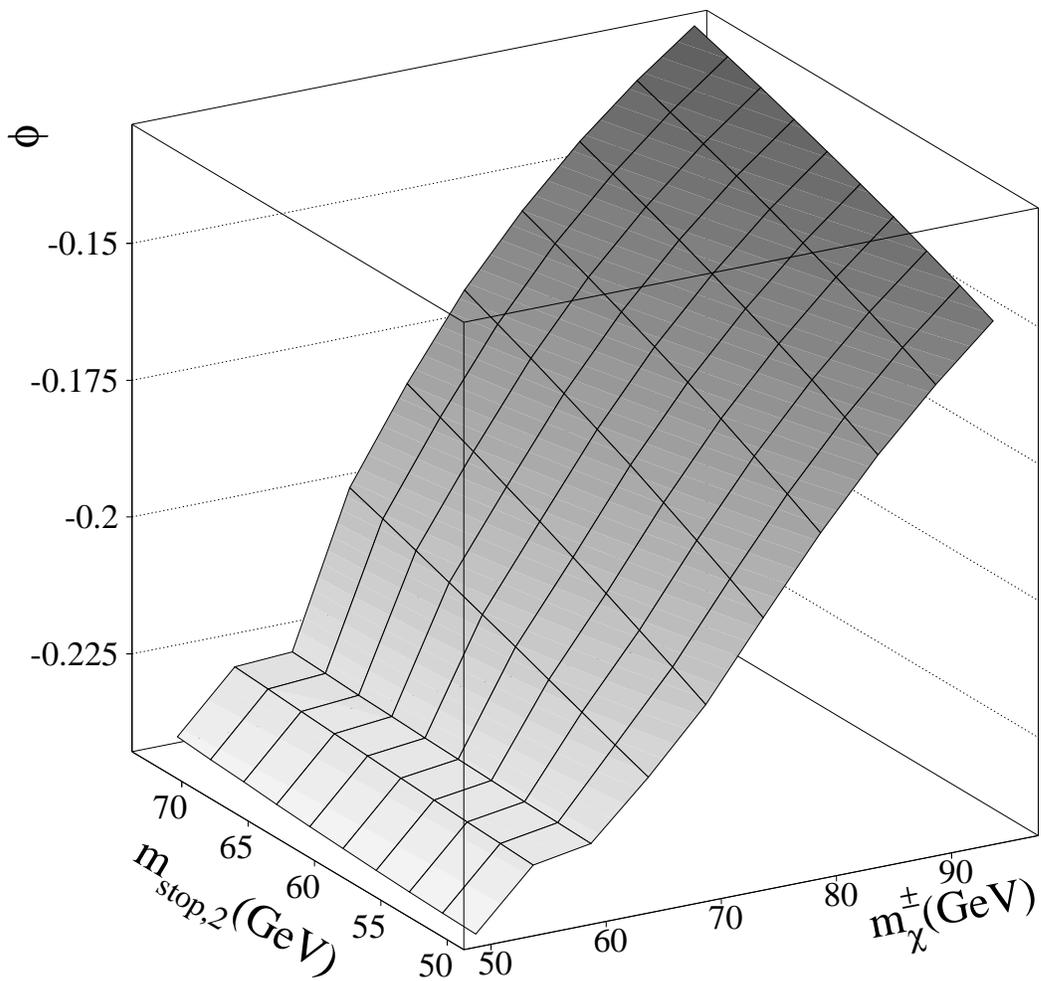}
 \end{center}
\caption{\label{\figomix}
Stop mixing angle $\phi_{mix}$ in the  light stop and light
chargino plane. It is mainly determined by the $R_{b\rightarrow s\gamma}$ rate.
 Optimization as in  fig.~\protect\ref{\figochi}.
} 
\end{figure*}

\begin{figure*}
 \begin{center}
  \leavevmode
  \epsfxsize=15cm
  \epsffile{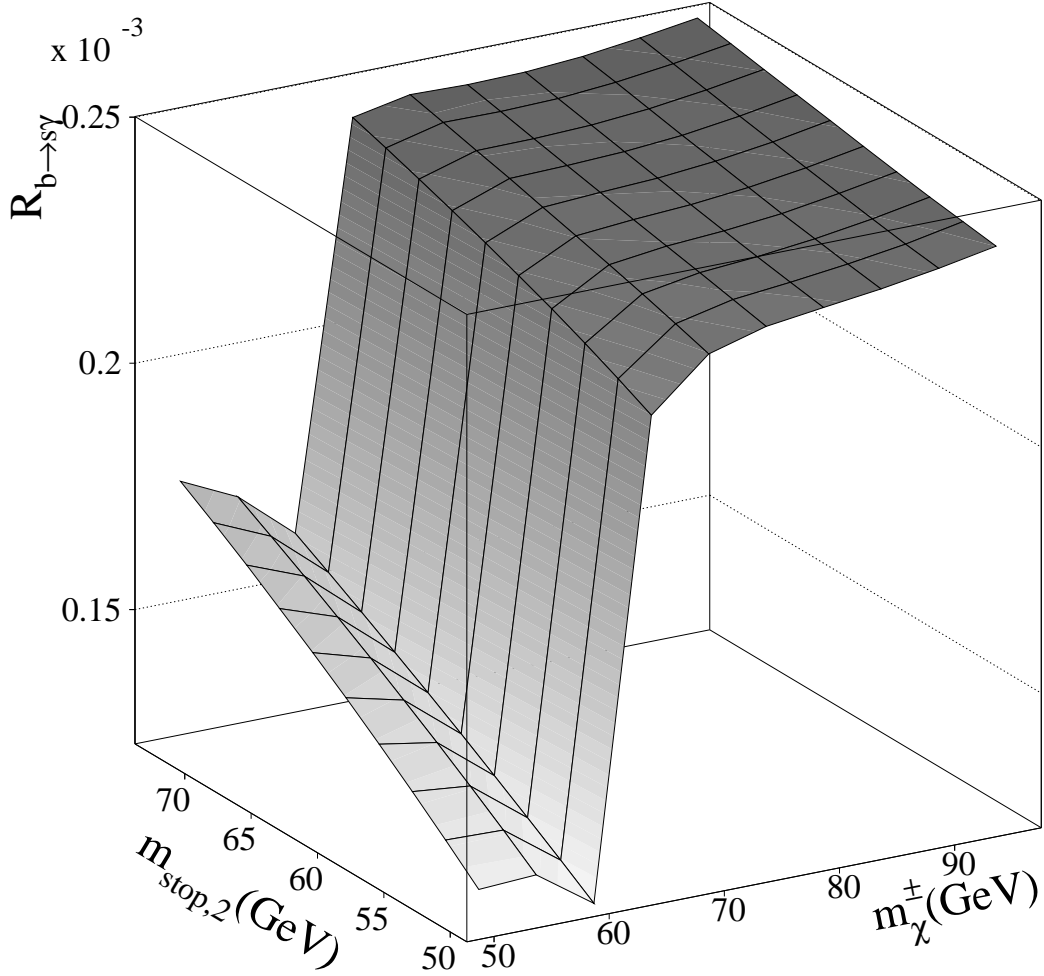}
 \end{center}
\caption{\label{\figobsg}
$R_{b\rightarrow s\gamma}$ in the  light stop and light
chargino plane for $\tan\beta=1.6$. For chargino masses
higher than 60~GeV (and $\mu > 0 $) the predicted
value is close to 
the CLEO measurement of
$2.32\pm0.67\times 10^{-4}$.
Optimization as in  fig.~\protect\ref{\figochi}.
} 
\end{figure*}

Fig.~\ref{\figochi} shows the change in the best obtainable $\chi^2$ in the chargino - stop plane.
For each value of the lighter scalar top $m_{\tilde t_{2}}$ and lighter chargino $m_{\tilde \chi_2^\pm}$
in a grid of 10$\times$10 points
an optimization of $m_t$, $\alpha_s$ and the stop mixing angle $\Phi_{mix}$ was performed,
assuming $M_2=3|\mu|$ for a fixed value of $\tan\beta=$1.6. 
In the next section this assumption on $M_2$ will be dropped.
Low sparticle masses yield a
sharp increase in the $\Delta\chi^2$ in fig.~\ref{\figochi} because of the included mass limits.
The minimum  $\chi^2$  is
obtained for chargino masses above 80~GeV and increases only slowly with
increasing sparticle masses.
$R_b$ increases significantly with decreasing values of
the stop and chargino mass, as can be seen from fig.~\ref{\figorb}.
Much less significant is the change of
$R_c$. Within the plane of fig.~\ref{\figorb} it changes less than 0.0005 units.
The increase of $R_b$ must be compensated by a decrease  of $\alpha_s$ (see fig.~\ref{\figoalf})
in order to keep the total $Z^0$-width constant.
The stop mixing angle $\Phi_{mix}$, shown in fig.~\ref{\figomix}, is mainly determined by the CLEO
measurement of
$R_{b\rightarrow s\gamma}$.
The chargino contribution to $R_{b\rightarrow s\gamma}$ is proportional to the Higgs mixing parameter
$\mu$, which changes its sign for $m_{\tilde\chi^{\pm}}\approx$ 60~GeV (see fig.\ref{\figI}), so the
$R_{b\rightarrow s\gamma}$ rate changes rapidly for these chargino masses, as shown in  fig.~\ref{\figobsg}.

\subsection{High $\tan\beta$ scenario} 

\begin{figure*}
 \begin{center}
  \leavevmode
  \epsfxsize=15cm
  \epsffile{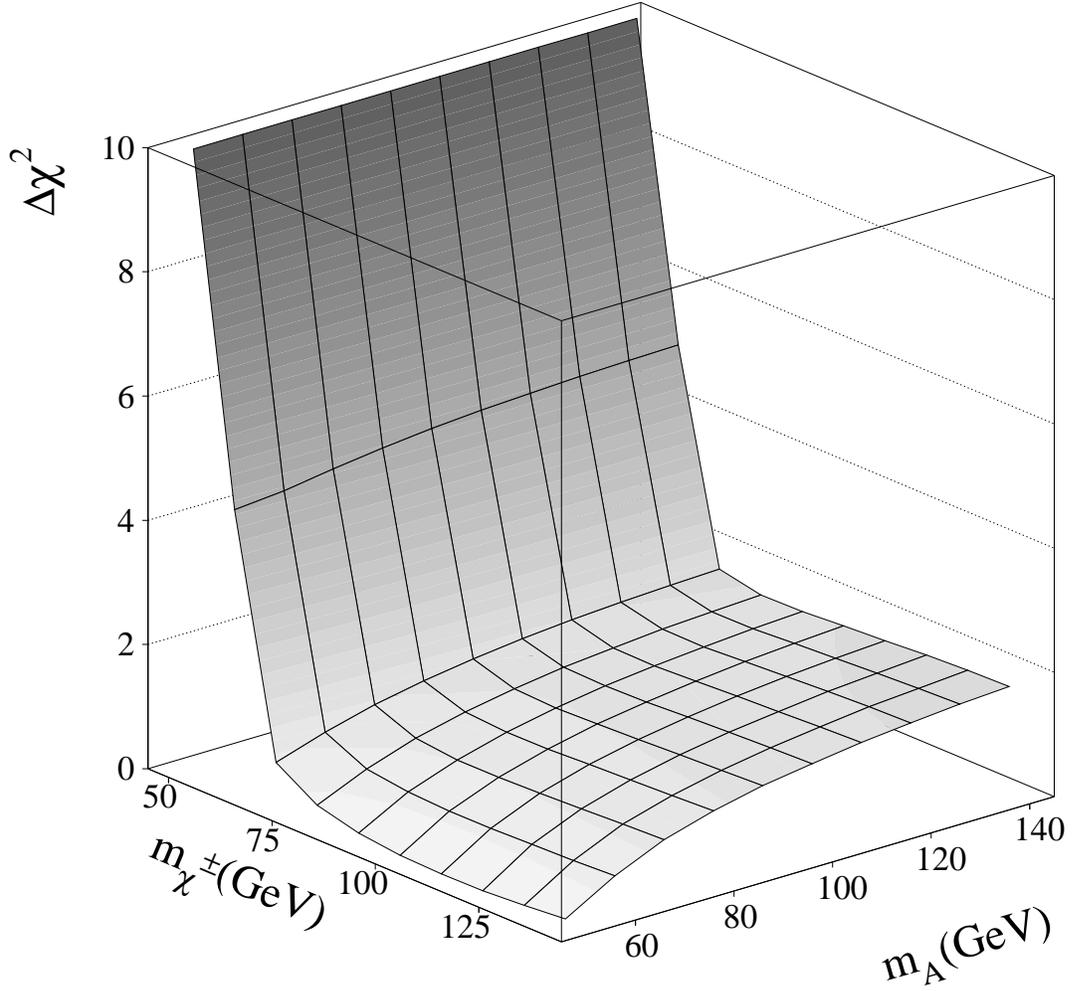}
 \end{center}
\caption{\label{\fighochi}
The $\Delta\chi^2$ in the pseudoscalar Higgs and light chargino plane for $\tan\beta=35$.
For each given $m_A$ and light 
chargino mass an optimization of $m_t$ , $\alpha_s$ , $m_{\tilde{t}_2}$ and the stop
mixing angle $\phi_{mix}$ was performed.
The irrelevant parameter $M_2$ was set to 1500~GeV. 
} 
\end{figure*}

\begin{figure*}
 \begin{center}
  \leavevmode
  \epsfxsize=15cm
  \epsffile{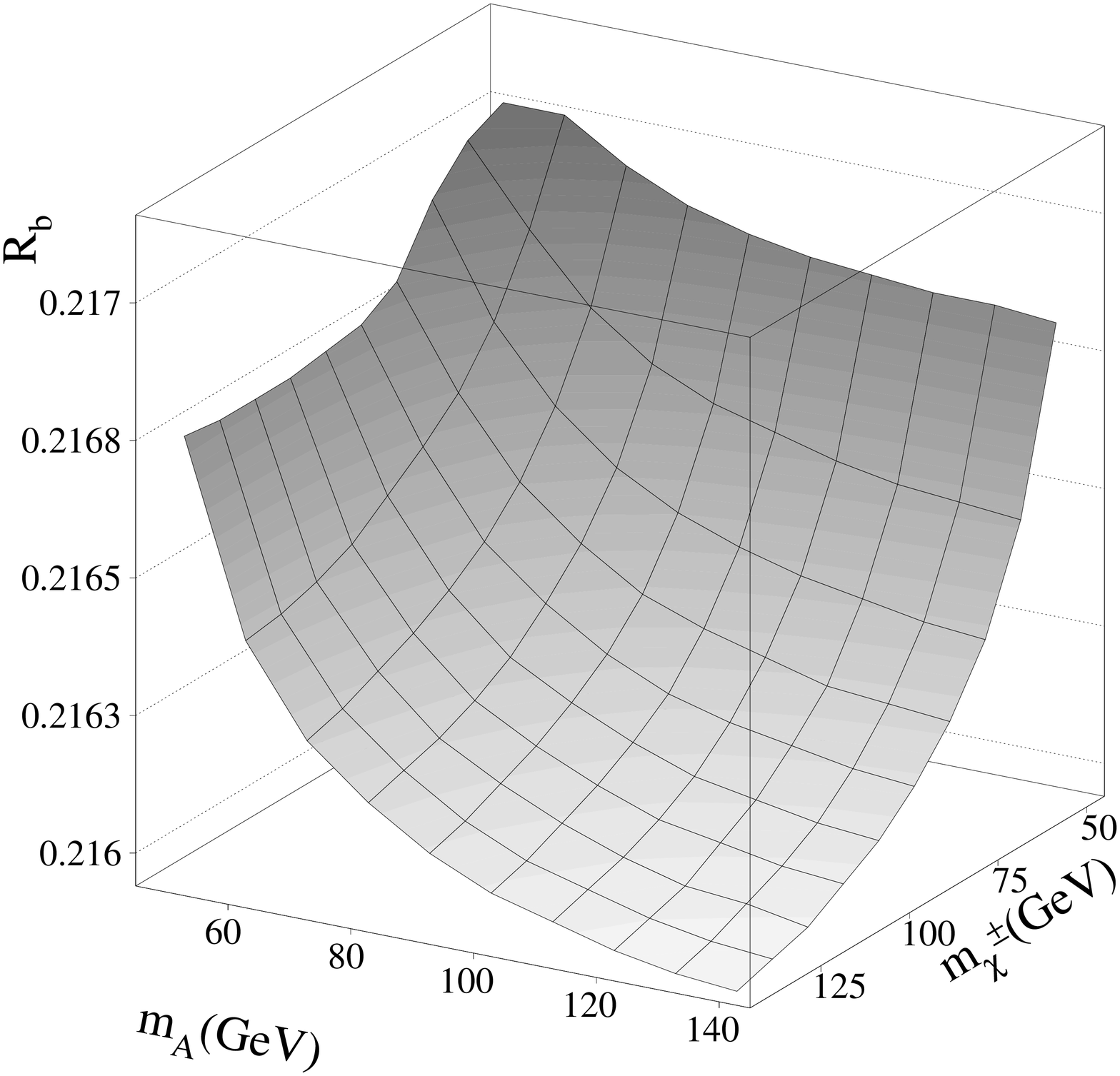}
 \end{center}
\caption{\label{\fighorb}
$R_b$ in the pseudoscalar Higgs and light chargino plane for $\tan\beta=35$.
 Optimization as in  fig.~\protect\ref{\fighochi}.
 } 
\end{figure*}

\begin{figure*}
 \begin{center}
  \leavevmode
  \epsfxsize=15cm
  \epsffile{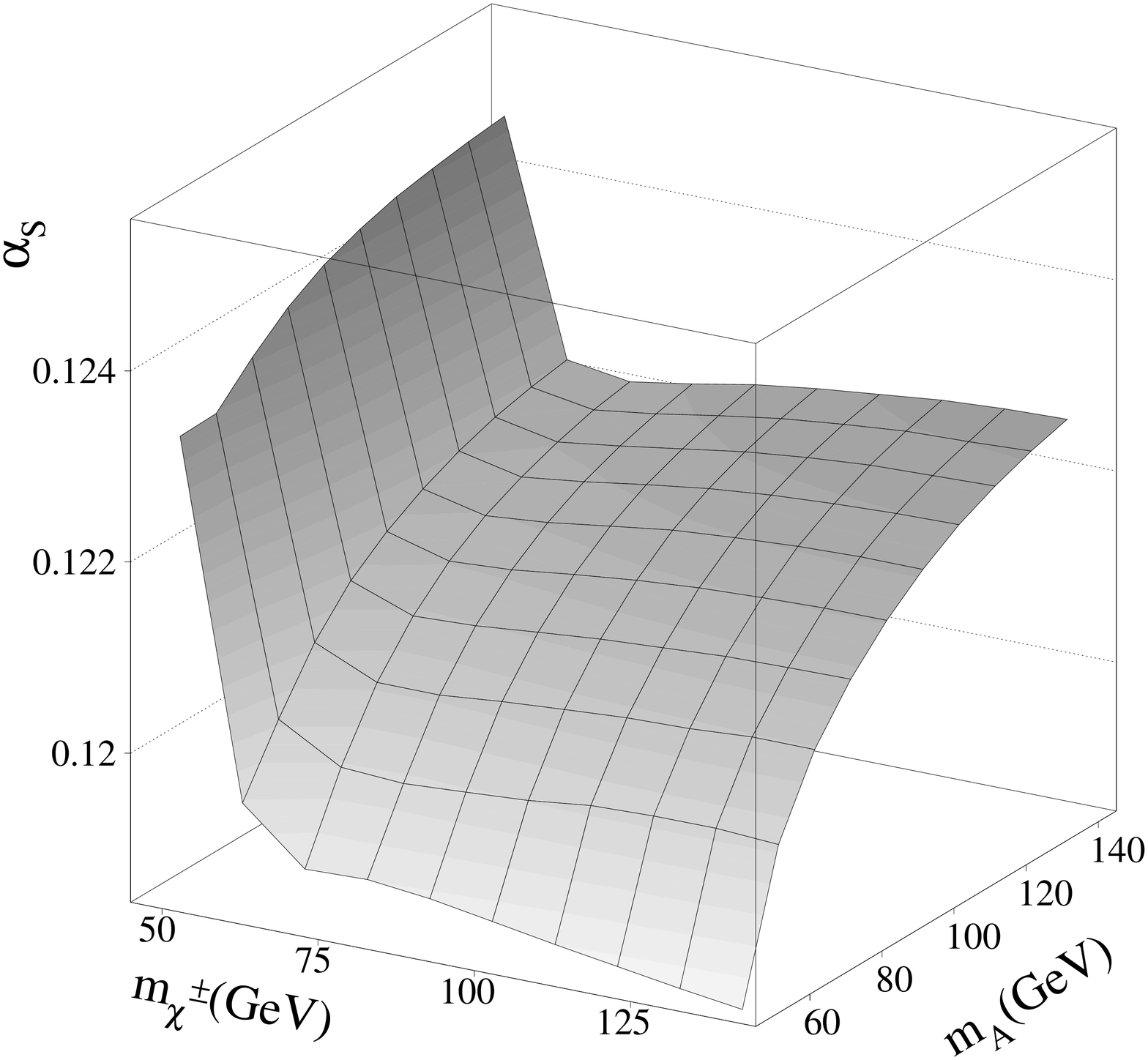}
 \end{center}
\caption{\label{\fighoalf}
$\alpha_s$ in the pseudoscalar Higgs and light chargino plane for $\tan\beta=35$.
Optimization as in  fig.~\protect\ref{\fighochi}.
} 
\end{figure*}

\begin{figure*}
 \begin{center}
  \leavevmode
  \epsfxsize=15cm
  \epsffile{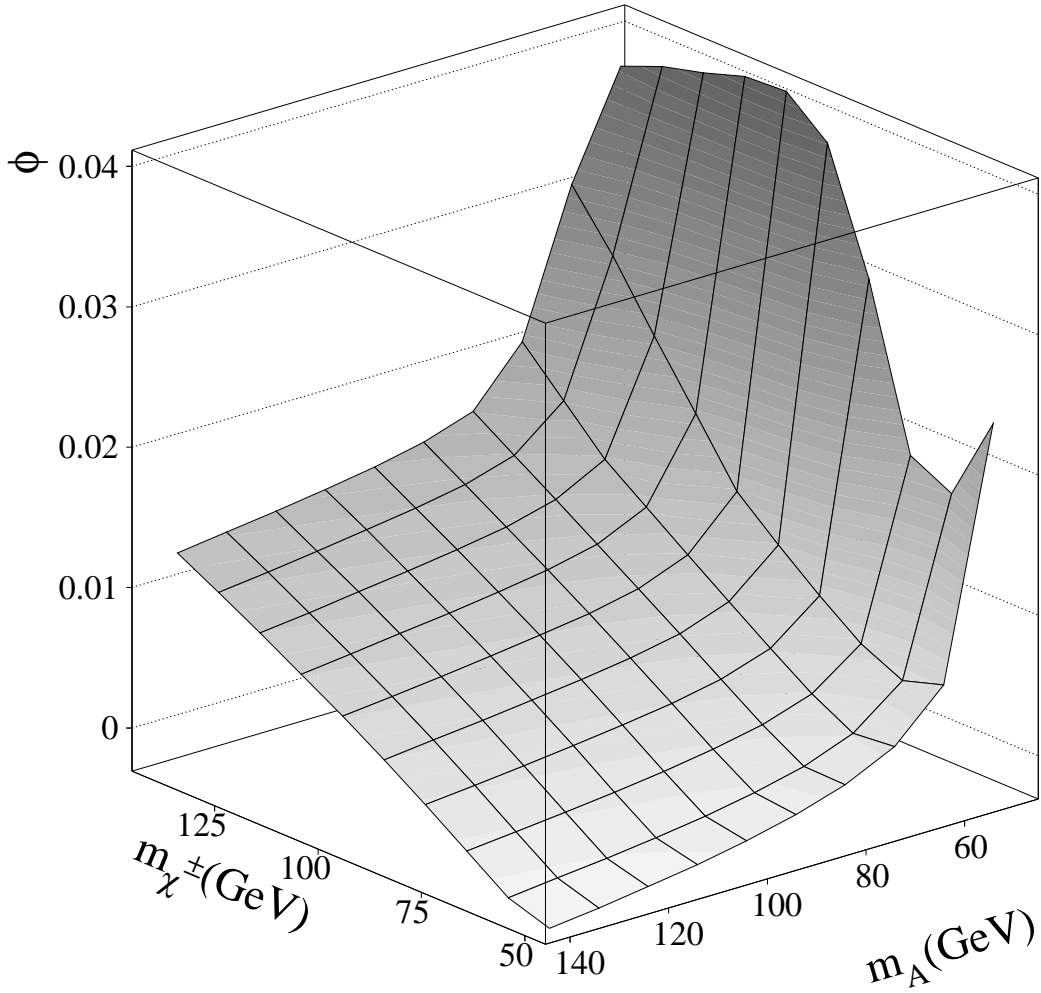}
 \end{center}
\caption{\label{\fighomix}
Stop mixing angle $\phi_{mix}$  in the pseudoscalar Higgs
and light chargino plane for $\tan\beta=35$.
Optimization as in  fig.~\protect\ref{\fighochi}.
} 
\end{figure*}

\begin{figure*}
 \begin{center}
  \leavevmode
  \epsfxsize=15cm
  \epsffile{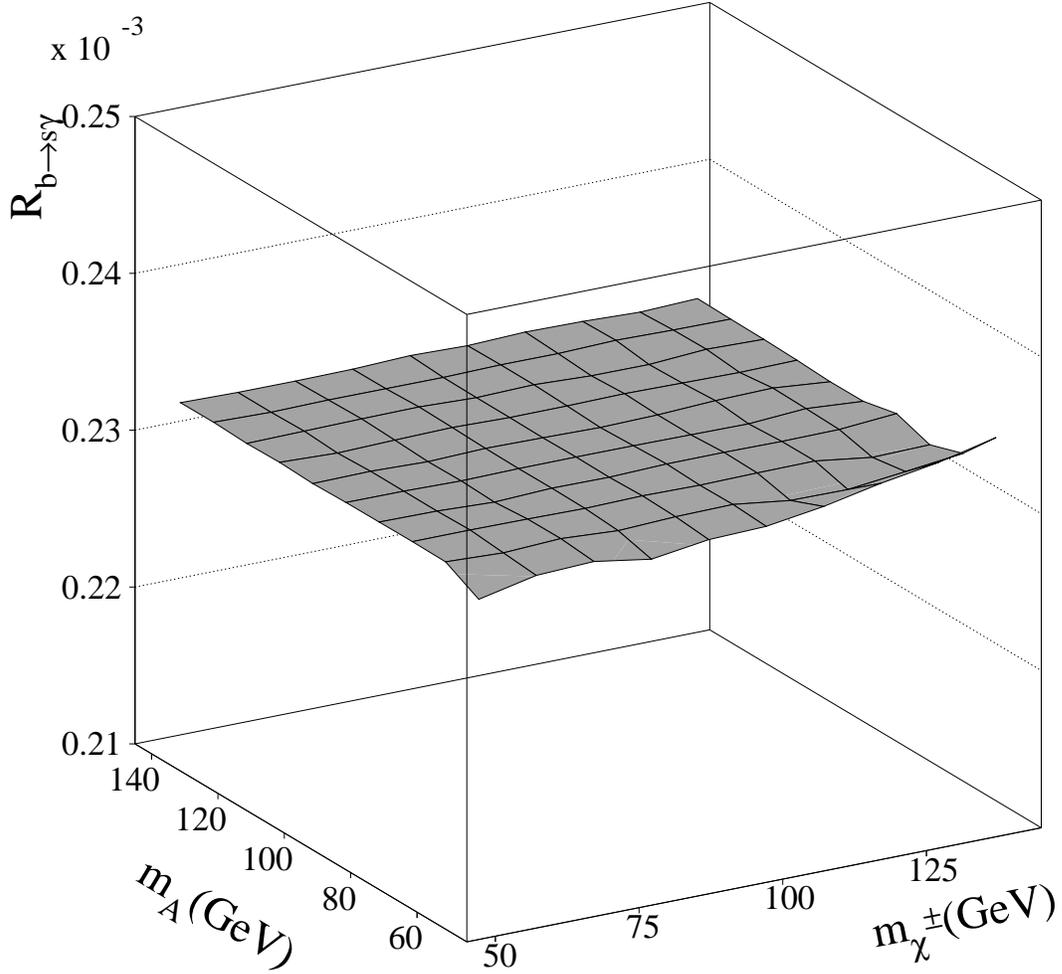}
 \end{center}
\caption{\label{\fighobsg}
$R_{b\rightarrow s\gamma}$ in the pseudoscalar Higgs
and light chargino plane for $\tan\beta=35$.
The prediction is close to the
CLEO measurement of 
$2.32\pm0.67\times 10^{-4}$
within the whole parameter space.
Optimization as in  fig.~\protect\ref{\fighochi}.
} 
\end{figure*}

Similar fits can be performed for the high $\tan\beta$ scenario  
in the pseudoscalar Higgs $m_A$ versus
light chargino plane.
In fig.~\ref{\fighochi}  the resulting change in the $\chi^2$ is
given for fixed $\tan\beta=35$. For small chargino masses there is 
a sharp increase in the
$\chi^2$ due to the corresponding mass limit, see above. 
The highest values for $R_b$ can be
obtained for small values of $m_A$ and $m_{\tilde\chi^{\pm}}$, 
see fig.~\ref{\fighorb}.
As in the low $\tan\beta$ case
the enhancement of $R_b$ must be compensated by a decrease of $\alpha_s$,
see fig.~\ref{\fighoalf}, and
the change of $R_c$ is small, less than 0.0006 within the given
parameter plane. 
The mixing angle, shown in fig.~\ref{\fighomix}, is mainly 
determined by the $R_{b\rightarrow s\gamma}$ rate,
which can be fitted in the whole  $m_A$-$m_{\tilde{\chi}^\pm}$ plane, 
see fig.~\ref{\fighobsg}.

\subsection{Best Solutions}

{\it Standard Model Fits:}\\
\vspace{0.1cm}

\begin{figure*}
 \begin{center}
  \leavevmode
  \epsfxsize=15cm
  \epsffile{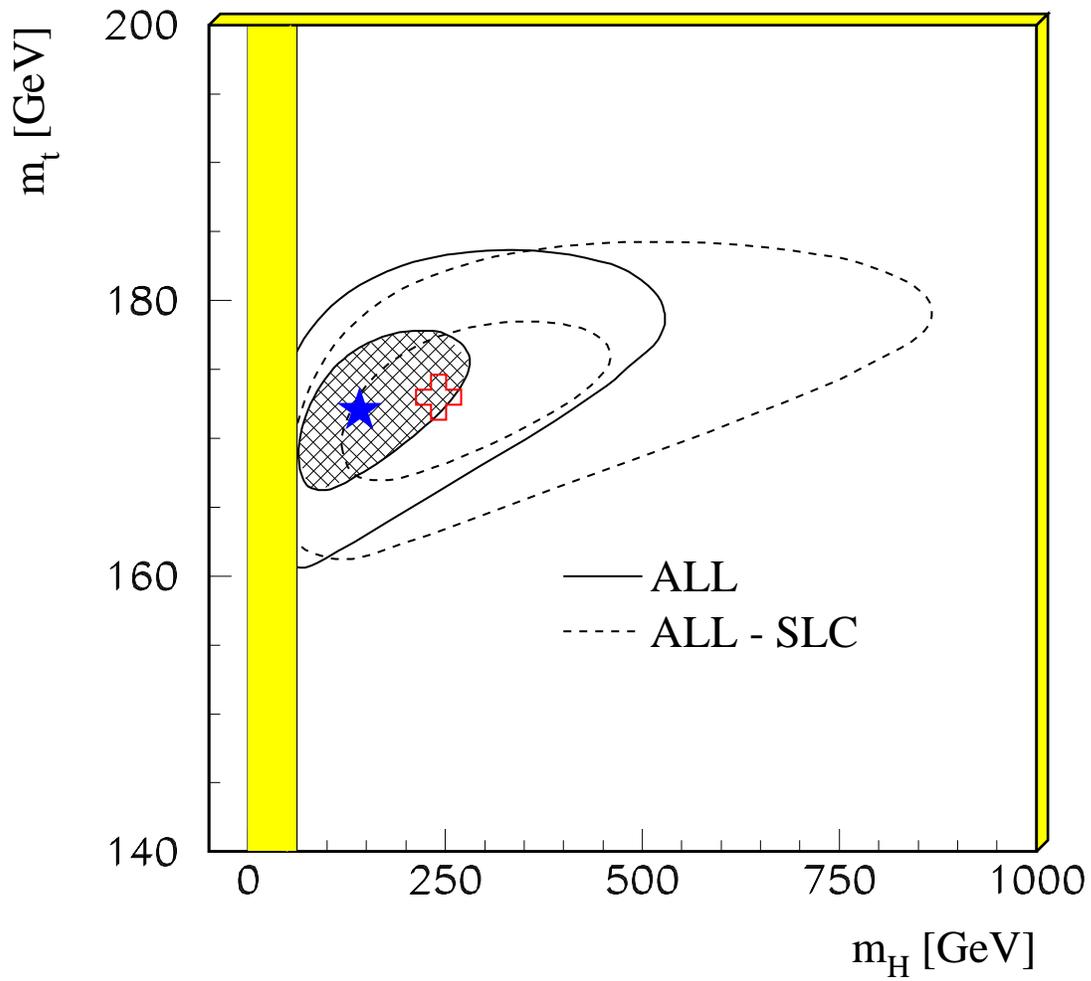}
 \end{center}
\caption{\label{\mtmh}
$\Delta\chi^2=1$ and $\Delta\chi^2=4$ contour lines for all electroweak data
including  $\sin^2\theta_{eff}^{lept}$ from SLD (continous line) and without it
(dashed line). The stars indicate the best fits.
} 
\end{figure*}

\begin{figure*}
 \begin{center}
  \leavevmode
  \epsfxsize=15cm
  \epsffile{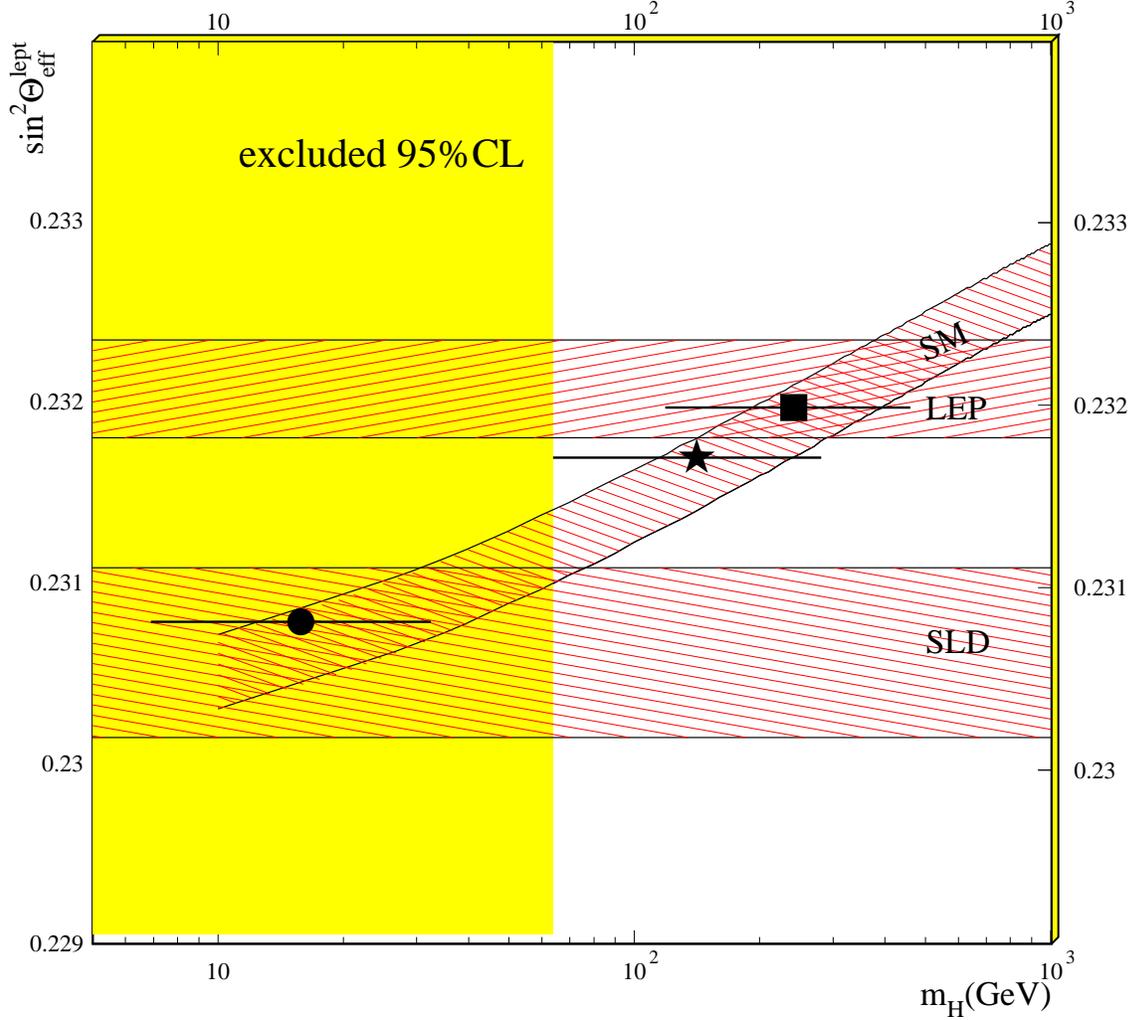}
 \end{center}
\caption{\label{\sintw}
Dependence of the SM $\sin^2\theta_{eff}^{lept}$ on the Higgs mass. The top
mass $m_t=175\pm 9$~GeV was varied within its error, as shown by the dashed band
labelled SM (upper (lower) boundary $m_t$=166(184)~GeV). 
The SLD and the LEP measurement of  $\sin^2\theta_{eff}^{lept}$ 
are also shown as horizontal bands.
The star and the square are respectivly the results of the combined fit to SLD and LEP data and LEP data without
the SLD measurement of $\sin^2\theta_{eff}^{lept}$, whereas the circle indicates the Higgs mass corresponding
to the SLD measurement of $\sin^2\theta_{eff}^{lept}$. 
Clearly,
the SLD value yields a Higgs mass less than the combined LEP limit of 58.4~GeV (shaded area).} 
\end{figure*}

The predictions of the SM are completely determined by the set of
six input parameters $M_Z$, $m_t$, the SM Higgs mass $M_h$,
$\alpha_s$, $\alpha_{EM}$ and $G_\mu$. The error of the muon decay
constant is so small that $G_\mu$ can be treated as a fixed parameter.
The fine structure constant was taken to be
$1/\alpha(M_Z)=128.89\pm0.09$\cite{jegerlehner}.
The error on $1/\alpha(M_Z)$ turns out not to be negligible: fixing
$1/\alpha(M_Z)$ underestimates the error on the Higgs mass
by $\approx 30$\%.
The SM predictions were obtained from the ZFITTER
package \cite{zfitter} and all the error correlations were taken from \cite{Lep3}.
The fits were made with  $M_Z$, $m_t$,
$m_H$, $\alpha_s$ and $\alpha$ as free parameters, which resulted in
\begin{eqnarray*}
M_Z & = & 91.186\pm0.002~{\rm GeV}\\
m_t & = & 172.0^{+5.8}_{-5.7}~{\rm GeV}\\
m_H & = &141^{+140}_{-77}~{\rm GeV}\\
\alpha_s(M_Z) & = & 0.1197\pm0.0031\\
1/\alpha(M_Z) & = & 128.90^{+0.089}_{-0.090}\\
\end{eqnarray*}
%
The minor differences to the results given in
\cite{Lep3} originate from
the inclusion of the $R_{b\rightarrow s\gamma}$ ratio.
The electroweak mixing angle can be determined from these parameters:
\begin{eqnarray*}
\sin^2\theta_{\overline{MS}}=0.2316\pm0.0004.
\end{eqnarray*}
Both the strong coupling constant and the electroweak mixing angle 
$\sin^2\theta_{\overline{MS}}$ 
have been determined at the  scale $M_Z$ in the $\overline{MS}$
renormalization scheme. With $m_t=175$ GeV $\sin^2\theta_{\overline{MS}}=
 \sin ^2\theta_{eff}^{lept}$ to an accuracy better than 0.0001.
The quoted errors have been determined using MINOS~\cite{minuit}.
Further details of the  procedure are described  elsewhere, 
see for example~\cite{ralf,cmssm}.
The $\chi^2/d.o.f$ of
the SM fit is 19.6/15 which corresponds to a probability of 19\%. Here, the main contributions to the
$\chi^2$ originate from $\sin^2\theta_{eff}^{lept}$ from SLD  ($\Delta\chi^2=4.9$), $R_b$ ($\Delta\chi^2=3.1$) 
and $A_{FB}^b$  ($\Delta\chi^2=3.5$). The prediction of $R_c$ is good.
If $R_{b\rightarrow s\gamma}$ is excluded from the fit, one obtaines a $\chi^2/d.o.f$ of 18.9/14, corresponding to
a probability of 17\%, in agreement with \cite{Lep3}.
The correlation parameter
between $m_H$ and $m_t$ for the best fit is approximately 0.7; this strong correlation is shown in fig. \ref{\mtmh}.
One observes that the upper limit on the Higgs mass is obtained for $m_t$ above $175$~GeV; however, the
upper limit is sensitive to $\sin^2\theta_{eff}^{lept}$ as shown by the dashed contour in fig.~\ref{\mtmh}, where the
precise value of $\sin^2\theta_{eff}^{lept}$ from SLD was excluded from the fit.
The  dependence of $\sin^2\theta_{eff}^{lept}$ on the SM Higgs mass is approximately logarithmic
(see fig. \ref{\sintw}).
The LEP data  without SLD yields $m_H=241^{+218}_{-123}$~GeV, while the SLD value of  $\sin^2\theta_{eff}^{lept}$
corresponds to $m_H=16^{+16}_{-9}$~GeV; both $m_h$ values are
 indicated in  figure \ref{\sintw} (by the square and the circle, respectively). 
The SLD value is excluded by the lower limit of 58.4~GeV from the combined LEP experiments~\cite{rev96}, so more data is eagerly awaited.

\begin{figure*}
 \begin{center}
  \leavevmode
  \epsfxsize=15cm
  \epsffile{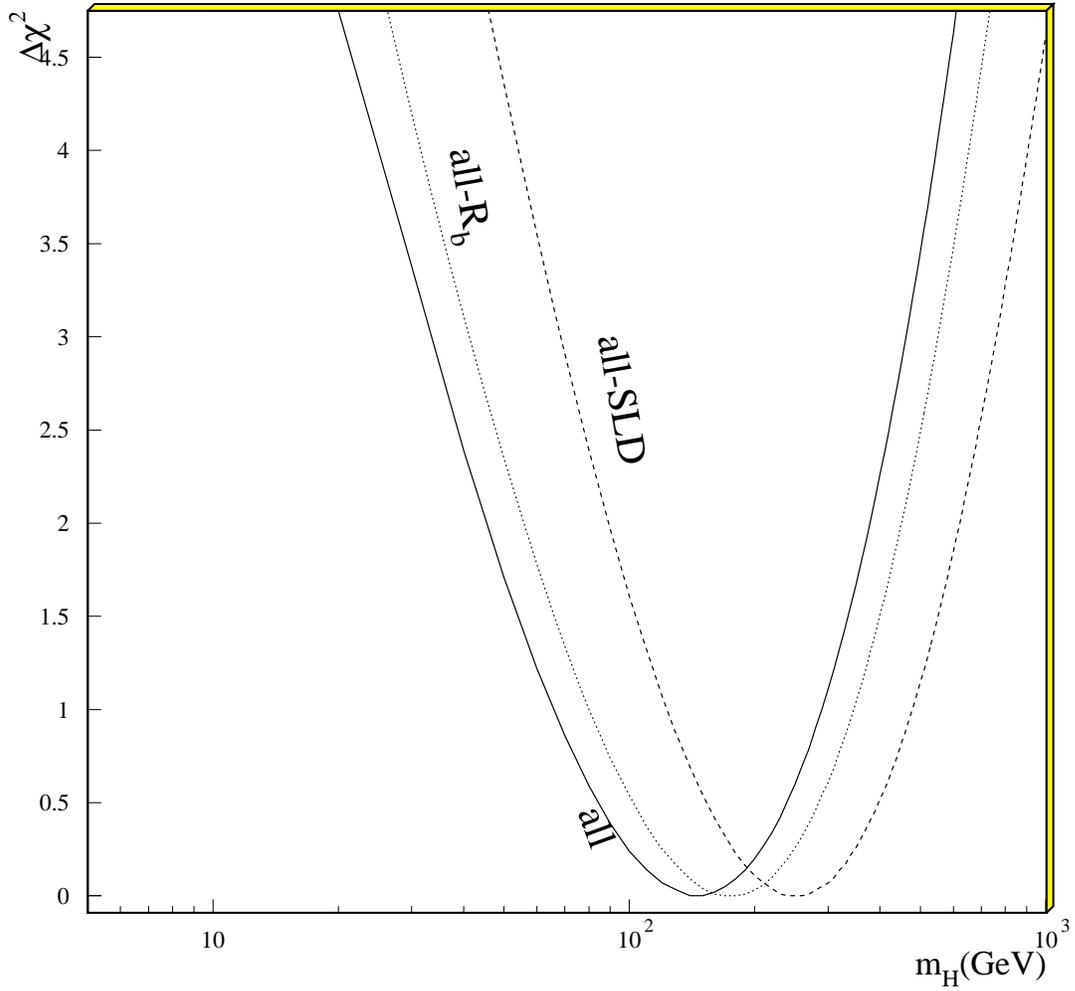}
 \end{center}
\caption{\label{\smdchi}
Dependence of the SM $\Delta\chi^2$ on the Higgs mass for a free top mass,
taking all data (continous line), all data
without the SLD measurement of  $\sin^2\theta_{eff}^{lept}$ (dashed line) and
all data without $R_b$ (dotted line). 
} 
\end{figure*}

The $\Delta\chi^2$ dependence of the Higgs mass is shown in fig. \ref{\smdchi} for various conditions.\\
Clearly, the $\sin^2\theta_{eff}^{lept}$ from SLD gives a large weight, while  the new value from $R_b$
plays only in minor role in contrast to the previous value\cite{Lep2}. 
\vspace{0.5cm}

{\it MSSM Fits and Comparison with the SM:}\\

In order to obtain the best MSSM fits the assumption $M_2=3|\mu|$ is dropped
and $M_2$ is treated as a free parameter.
As discussed in section \ref{zeff} the dominant contributions vary for the high and
the low $\tan\beta$ scenario. The preferred  $\tan\beta$ values for these scenarios are
around 35 and 1.6, respectively. Since the fit is not very
sensitive to
the precise $\tan\beta$ value, it was fixed to these values.  
The fitted  MSSM parameters and the corresponding SUSY masses are given in table
\ref{bestfit};  the predicted values of all observables
and their pulls are
summarized in table \ref{fitresults}.
For the MSSM $M_Z$, $G_\mu$ and $\alpha$ were treated as fixed parameters 
because of their small errors compared with the uncertainties from the other
parameters.
The MSSM prediction of the W-boson mass is always higher
than the SM one, but the values of the strong coupling constant, the electroweak mixing angle
and the top mass are very similar in the MSSM:   
\begin{eqnarray*}
\alpha_s(M_Z)=0.116\pm0.005\\
m_t=172\pm 5~{\rm GeV}\\
\sin^2\theta_{\overline{MS}}=0.2315\pm0.0004. 
\end{eqnarray*}
These values are for the low $\tan\beta$ scenario, but for high $\tan\beta$
the same values are obtained, except for $\alpha_s=0.119\pm0.005$ in that case.
The remaining parameters are given in table \ref{bestfit}.
A direct comparison to the SM fits is given  in
figs.~\ref{\figV}-\ref{\figVI}. The  SM fit to the 20 measurements
of table \ref{fitresults} with 5 parameters yields
$\chi^2/d.o.f.=19.6/15$  which corresponds to a probability of 19\%,
while the MSSM fits
correspond to probabilities of 17\% ($\tan\beta=1.6$, $\chi^2/d.o.f.=16.6/12$) and 11\%
($\tan\beta$=35, $\chi^2/d.o.f.=18.1/12$).
In counting the d.o.f the insensitive (and fixed) parameters are ignored, so
only the parameters given in table \ref{bestfit} are considered.
The difference in $\chi^2$ between SM and MSSM fits is mainly caused by $R_b$, which is better described in the MSSM, although the difference  in $\chi^2$
is insufficient to distinguish between the models.

The Higgs mass is not an independent parameter in the MSSM, since the couplings in the Higgs potential
are gauge couplings, which limit the mass of the lightest Higgs to a rather narrow range\cite{wimhig}.
The high $\tan\beta$ needs a light   pseudoscalar Higgs mass. As the lightest Higgs mass is
strongly correlated with the pseudoscalar Higgs mass, it is also low.
Similar Higgs values in the MSSM model  were
obtained in ref.~\cite{ellisfogli}.

An interesting point is the fitted value of  $\alpha_s(M_Z)$.
In previous analysis with the high values of $R_b$, $\alpha_s(M_Z)$ in the MSSM
 ($\approx 0.11$)
was always significantly smaller than the SM value ($0.123\pm0.005$\cite{rev96}),
 which supported the low energy values from deep inelastic scattering (DIS)
 ($0.112\pm0.005$\cite{rev96})
and lattice calculations of the heavy quark 
splittings ($0.110\pm0.006$\cite{rev96}).
However, the discrepancies between the low energy $\alpha_s$ values and the LEP data
have practically disappeared
 at the Warsaw Conference\cite{schmelling}:
 the Standard Model value ($0.120\pm0.003$, see above) 
 is now in agreement with DIS measurements  
($0.115\pm0.005$\cite{schmelling}), lattice calculations $0.117\pm0.003\cite{schmelling}$
and the world average $0.118\pm0.003$\cite{schmelling}. The  MSSM values of 
$ \alpha_s$ are in
good agreement with these other determinations, as shown in fig. \ref{\alphas}.

The particle spectrum for the best fits, as shown in table \ref{bestfit},
suggests that some SUSY particles could be within  reach of LEP II.
Unfortunately, if the
 stop-, chargino- and/or Higgs mass are well above the discovery reach of   LEP II,
 the $\chi^2$ of the fit increases at most up to the SM value, since these
 particles basically decouple as soon as they become heavier than the heaviest SM
 mass, say the top mass of about 200 GeV. So one cannot get upper limits on these particles,
 since the probability changes only a few percent between the  SM and MSSM.

\section{CMSSM and $R_b$}
In the MSSM fits discussed above the lightest stop is mainly right-handed,
while the left-handed stop has to be heavy.
If both would be light, then all other squarks would  likely  be light, 
which would upset the good agreement between the SM 
and all other electroweak data.
A large mass splitting in the stop sector 
needs a very artificial fine tuning of the few free parameters in 
the Constrained MSSM, which assumes  unification of gauge and
b-$\tau$ Yukawa couplings\cite{cmssm}. 
This is obvious from the sfermion mixing matrix for the stop quarks,
eq.~\ref{sqmatrix}.
The D-terms proportional to $\cos 2\beta$ are negligible 
for $\tan\beta\approx 1$.
If one of the diagonal elements is much larger than $m_t$, the off-diagonal
terms of the order  $m_t$ will not cause a mixing and 
the difference between the left- and 
right-handed stops has to come from the evolution of the diagonal terms 
(for the notation see ref. \cite{cmssm}): 
\begin{eqnarray}
\small
\frac{dM^2_{\tilde{Q}_3}}{dt} & = & (\frac{16}{3}\tilde{\alpha}_3M^2_3
    + 3\tilde{\alpha}_2M^2_2 + \frac{1}{15}\tilde{\alpha}_1M^2_1)
      \label{RGEmQ} \\
 && - [Y_t(M^2_{\tilde{Q}_3}+M^2_{\tilde{U}_3}+m^2_{H_2}+A^2_t m_0^2)\nn
 &&    +Y_b(M^2_{\tilde{Q}_3}+M^2_{\tilde{D}_3}+m^2_{H_1}+A^2_b m_0^2)]\nn
\frac{dM^2_{\tilde{U}_3}}{dt} & = & (\frac{16}{3}\tilde{\alpha}_3M^2_3
    +\frac{16}{15}\tilde{\alpha}_1M^2_1)\\
&&    -2Y_t(M^2_{\tilde{Q}_3}+M^2_{\tilde{U}_3}  
 + m^2_{H_2}+A^2_t m_0^2)\nonumber \label{RGEmU}
\label{RGEmD}
\end{eqnarray}
\normalsize
One observes that
the difference between left- and right handed stops, 
denoted by $\tilde{Q}_3$ and $\tilde{U}_3$, respectively), depends
on the Yukawa couplings for top and bottom ($Y_t,Y_b$) 
and the trilinear couplings $A_{t(b)}$. For low $\tan\beta$
$Y_b$ is negligible, while  $A_t$ and $Y_t$ are not free parameters, 
since they  go to 
 fixed point solutions\cite{cmssm}. Therefore there is little freedom to
adjust these parameters within the CMSSM in order to get a large splitting
between the left- and right-handed stops.  
\normalsize

In addition, problems arise with electroweak symmetry breaking,
since this requires the Higgs mixing parameter $\mu$ 
to be much heavier than the gaugino masses\cite{cmssm}, 
while $R_b$ requires low values of $\mu$ for a 
significant enhancement (since the
chargino has to be preferably Higgsino-like).
In conclusion, within the CMSSM an enhancement of $R_b$ above 
the SM is practically excluded.

%

\section{Conclusions}
Both the MSSM and SM provide a good description of all electroweak data.
The best
$\chi^2/d.o.f$ in the MSSM (SM) is  16.6/12 (19.6/15), which  corresponds to a
probability of 17\% (19\%).
The lower $\chi^2$ of the MSSM is mainly 
due to the better description
of $R_b$, but the fit 
requires an unnatural large splitting in the stop sector, 
as discussed in the previous section.
  Since the final analysis of most of the available LEP data is still in progress,
one has to wait and see if the  
present preliminary value of $R_b$ will indeed stay above the SM prediction.

%

\clearpage

\vspace{10cm}

\clearpage

%
%

\clearpage
\newpage
\thispagestyle{empty}

\begin{figure*}
 \begin{center}
  \leavevmode
  \epsfxsize=15cm
  \epsffile{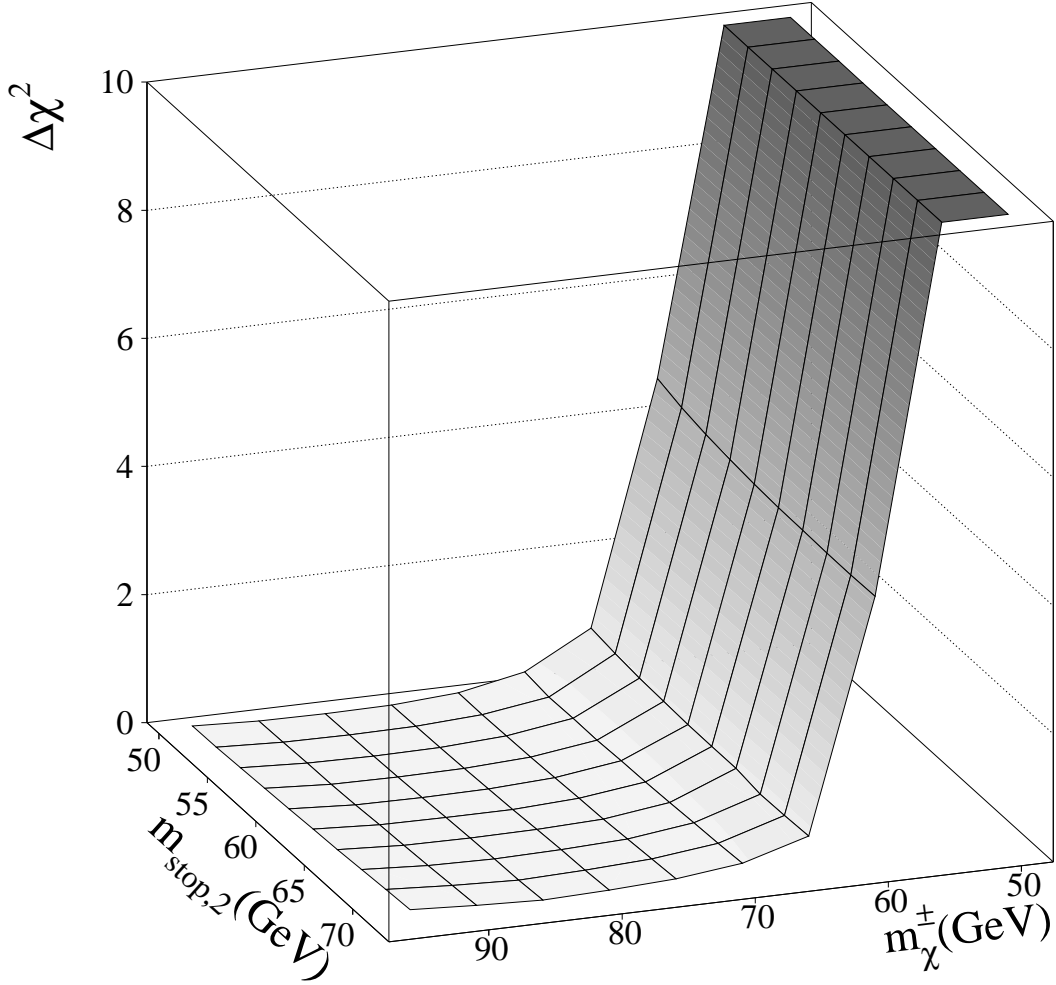}
 \end{center}
\caption{\label{\figochiall}
The $\Delta\chi^2$ in the region of the best fit in the  light stop and light
chargino plane for $\tan\beta=$1.6. Here the constraint on $M_2$ was dropped.
At each point of the grid an
optimization of $m_t$, $M_2$, $\alpha_s$
and the stop mixing angle $\phi_{mix}$ was performed with $\mu > -40$.
} 
\end{figure*}


\clearpage
\newpage
\thispagestyle{empty}
\begin{figure*}
 \begin{center}
  \leavevmode
  \epsfxsize=15cm
  \epsffile{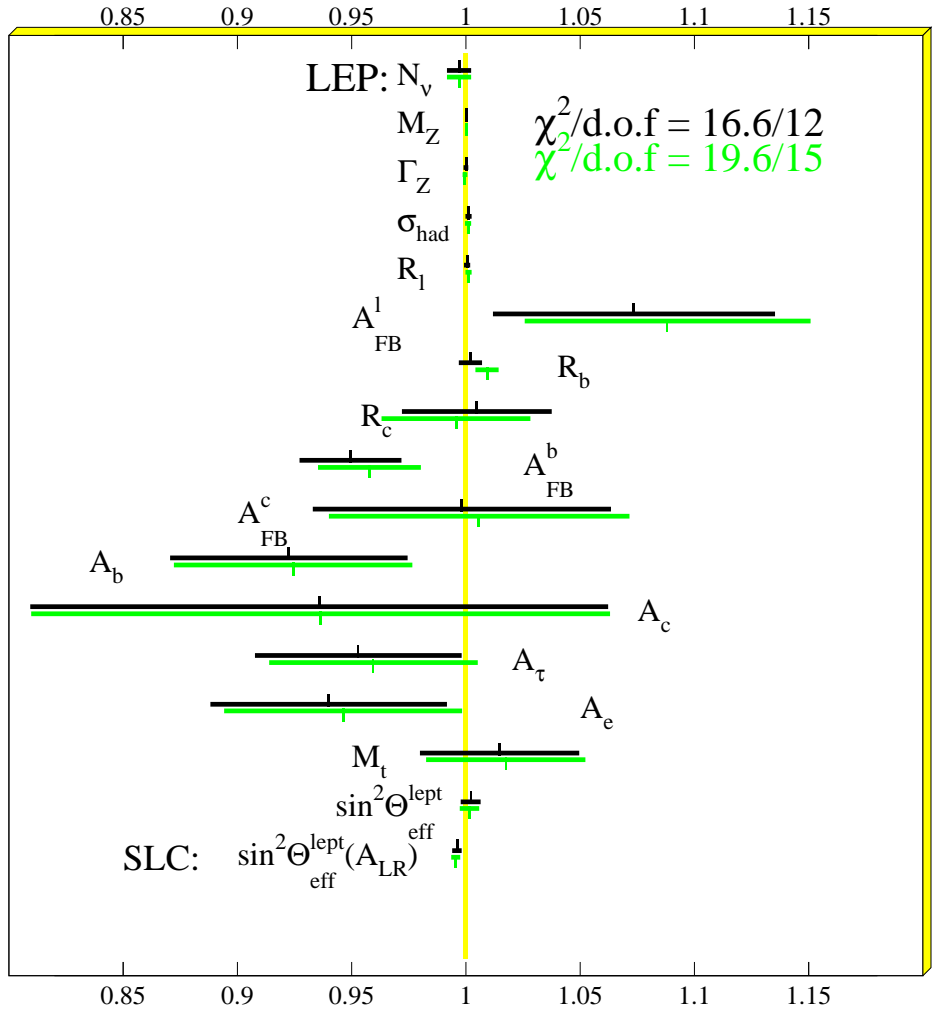}
 \end{center}
\caption{\label{\figV}
Resulting observables for the fit given in table \protect\ref{bestfit}
for $\tan\beta=1.6$. $m_{\tilde b}$ was
fixed to 1000~GeV, $m_A$ and the gluino mass were fixed to 1500~GeV.
The remaining deviation of $R_b$ from the SM prediction can be fully explained
within the MSSM.}
\end{figure*}

\clearpage
\newpage
\thispagestyle{empty}
\begin{figure*}
 \begin{center}
  \leavevmode
  \epsfxsize=15cm
  \epsffile{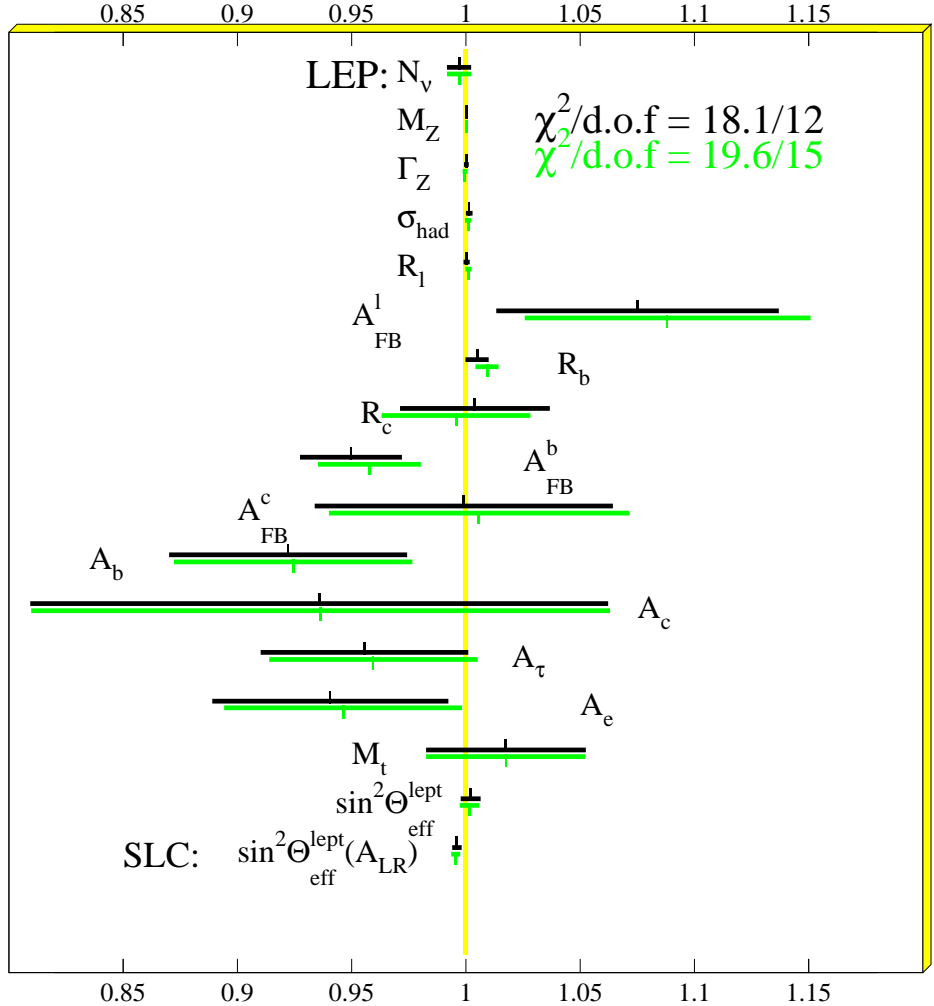}
 \end{center}
\caption{\label{\figVI}
Resulting observables for the fit given in table \protect\ref{bestfit}
for $\tan\beta=35$. $m_{\tilde b}$ was
fixed to 1000~GeV, $M_2$ and the gluino mass were fixed to 1500~GeV. It is possible
to improve the prediction of $R_b$ with Supersymmetry even for high values of
$\tan\beta$, but the result is not as good as for low values.}
\end{figure*}

\begin{figure*}
 \begin{center}
  \leavevmode
  \epsfxsize=14cm
  \epsffile{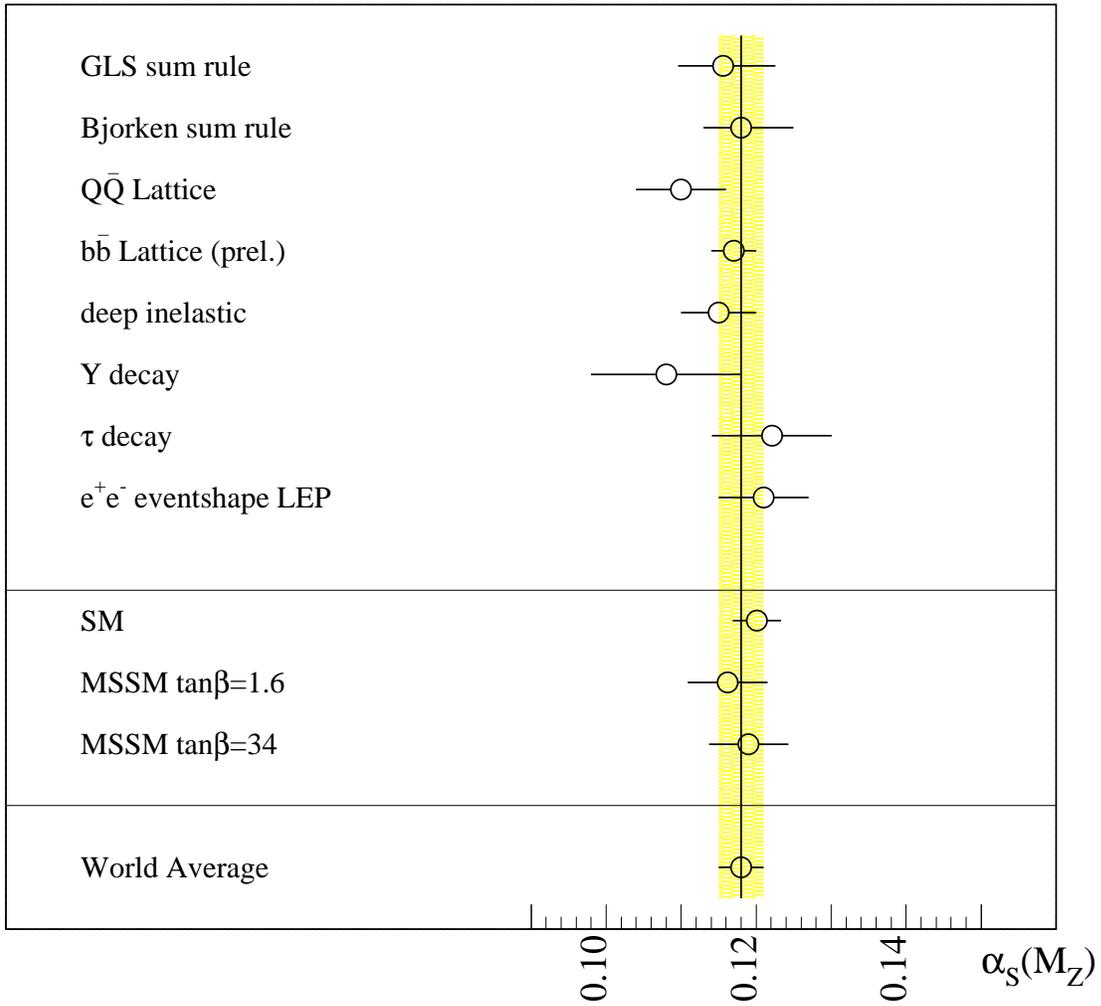}
 \end{center}
\caption{\label{\alphas}
Comparison of different measurements of $\alpha_s$ with our fit results, labeled
SM and MSSM.
The data has been taken from \protect\cite{rev96} and \protect\cite{schmelling}.}
\end{figure*}

\clearpage
\newpage
                                                           
%
\end{document}